%% file: paper_arxiv_v2.tex
\documentclass[transmag]{IEEEtran}

\IEEEoverridecommandlockouts

\usepackage{svg}
\usepackage{textcomp}
\usepackage{hyperref}
\usepackage{lipsum}
\usepackage{textcomp}
\usepackage{cite}
\usepackage{amsmath,amssymb,amsfonts}
\usepackage{algorithmic}
\usepackage{graphicx}
\usepackage{multicol,multirow}
\usepackage{array}
\usepackage{gensymb}
\usepackage{mathtools, cuted}
\usepackage{breqn}
\usepackage{lipsum}
\usepackage{xcolor}
\usepackage{siunitx}
\usepackage{subfloat}
\usepackage{subfig}

\usepackage{tikz}
\usepackage{tikzsymbols}
\usetikzlibrary{calc}
\usepackage[europeanresistors,americaninductors]{circuitikz}

\usepackage{pgfplots}
\usepackage{lipsum}

\usepackage{adjustbox}

\usepackage{import}
\usepackage{placeins}
\usepackage{calc}

\usepackage{dblfloatfix}

\usepackage[font=small]{caption}  

\newcolumntype{L}{>{\raggedright\arraybackslash}p}
\newcolumntype{C}{>{\centering\arraybackslash}m}

\ctikzset{resistors/scale=0.8, 
capacitors/scale=0.7, 
inductors/scale=0.7, 
transistors/scale=1.3} 

\begin{document}

\title{Electromagnetic Model of Reflective Intelligent Surfaces}

\author{Filippo Costa,~\IEEEmembership{Senior Member,~IEEE}
	and~Michele Borgese,~\IEEEmembership{Member,~IEEE}
	
	\thanks{Filippo Costa is with University of Pisa, Dipartimento di Ingegneria dell'Informazione, Pisa, Italy. (corresponding author, e-mail: filippo.costa@unipi.it)
	
	Michele Borgese is with the Research and Development Department, SIAE MICROELETTRONICA, 20093 Cologno Monzese, Italy. (e-mail: michele.borgese@siaemic.com)}
}

\onecolumn 
\begin{quote}
\begin{center}
F. Costa and M. Borgese, "Electromagnetic Model of Reflective Intelligent Surfaces," in \textit{IEEE Open Journal of the Communications Society}, vol. 2, pp. 1577-1589, 2021, doi: 10.1109/OJCOMS.2021.3092217.
\end{center}
\end{quote}

\twocolumn

\maketitle

\begin{abstract}
	\boldmath
	An accurate and simple analytical model for the computation of the reflection amplitude and phase of Reconfigurable Intelligent Surfaces is presented. The model is based on a transmission-line circuit representation of the RIS which takes into account the physics behind the structure including the effect of all relevant geometrical and electrical parameters. The proposed representation of the RIS allows to take into account the effect of incidence angle, mutual coupling among elements and the effect of the interaction of the periodic surface with the RIS ground plane. It is shown that, the proposed approach allows to design a physically realisable RIS without recurring to onerous electromagnetic simulations. The proposed model aims at filling the gap between RIS assisted communications algorithms and physical implementation issues and realistic performance of these surfaces.
\end{abstract}


\section{Introduction}
\label{sec_introduction}

Reconfigurable Intelligent surface (RIS) \footnote{Intelligent Reflecting Surface (IRS) is alternatively used in other works.} assisted wireless communications have recently emerged as a promising solution to enhance the spectrum and energy efficiency of future wireless systems \cite{rappaport2019wireless,wu2019towards,wu2019beamforming, hu2018beyond, pan2020intelligent}. Specifically, a RIS allows to control the wireless propagation environment via an array of reconfigurable passive reflecting elements \cite{abeywickrama2020intelligent, dardari2020communicating, di2020smart}. RIS technology is attractive from an energy consumption point of view since it is possible forwarding the incoming signal without employing power amplifiers like in MIMO arrays \cite{ngo2013energy, larsson2014massive, rocca2016unconventional}, but rather by suitably designing the phase shift applied by each reflecting element in order to constructively combine the reflected signal. RIS can be also employed for physical information encoding in the so called massive backscatter wireless communication (MBWC) scheme \cite{zhao2020metasurface}, for simultaneous Wireless Information and Power Transfer (SWIPT), spectrum sharing, Non-orthogonal multiple access (NOMA) \cite{rajatheva2020white}. RIS structures have the advantage of being easily integrable in the communication environment because of the easy deployment into buildings, ceilings of factories or into human clothing \cite{huang2019reconfigurable}. 
A RIS consists essentially of a periodic metasurface \cite{borgese2020simple} including capacitive elements located at a sub-wavelength distance from a metallic plane. The metasurface unit cells are loaded with active components in order to control the phase and the amplitude of the reflection coefficient. The unit cells can be tuned such that signals bouncing off a RIS are combined constructively to increase signal in the position of the intended receiver or destructively to avoid leaking signals to undesired receivers \cite{mursia2020risma}.
The modeling, analysis and design of RISs for application to wireless networks is a multidisciplinary research topic covering communication theory, computer science and electromagnetism. A major limitation of current research on RISs in wireless networks is the lack of accurate models that describe the reconfigurable meta-surfaces as a function of their EM properties. 

\begin{figure}
	\centering
	\includegraphics[width=0.9\linewidth]{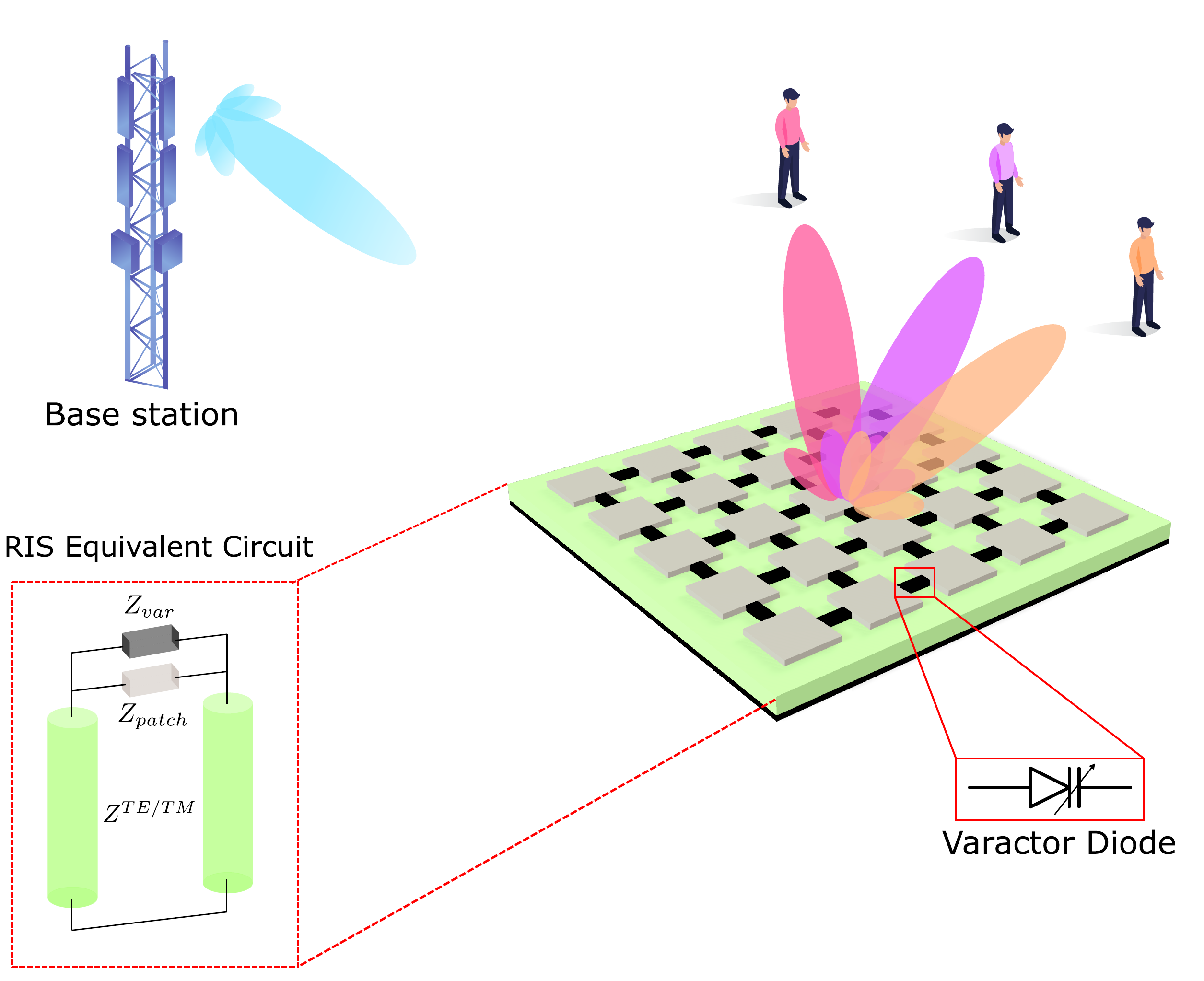}
	\hfill
	\caption{Three dimensional layout of the analyzed problem.}
	\label{fig_AIS} 
\end{figure}

The vast majority of research works available to date rely on the assumption that metasurfaces act as perfect reflectors with an arbitrarily controlled phase reflection. Some papers employ a fitting empirical model which is reasonable at normal incidence \cite{abeywickrama2020intelligent, najafi2020physics, tang2020mimo}. Actually, the response of RIS to the radio waves depends on the geometrical and electrical properties of the designed surface, on the choice of substrates (low-loss materials lead to lower losses but the cost of the surface scales rapidly with the size), the characteristics of the tunable components, but also on the properties of the impinging EM wave such as angle of incidence, angle of reflection \cite{chen2020angle}, polarization \cite{lukkonen_simple_model}. Physics based models which take into account the aforementioned design parameters are therefore of crucial importance in order to avoid unrealisable optimistic performance predictions \cite{basar2019wireless, alexandropoulos2020reconfigurable, najafi2020physics, tang2020mimo}. 
Moreover, designing RIS by neglecting some important aspects such as the change of the response at oblique incidence (spatial dispersion) does not allow to achieve optimal performance. The hard separation between communication and EM aspects is not fruitful as it would be important that communication algorithms could be supported by realistic, accurate and fast EM models. It is important that the EM counterpart of the problem is treated in real time with an accurate but still analytical approach since a high computational burden is already associated to the evaluation of RIS performance in complex multi-user the communication scenarios. As a consequence, a slow EM analysis including all the geometrical details of the periodic pattern cannot be acceptable . 
We present an accurate and simple analytical model for the computation of the amplitude and phase response of the RIS which takes into account incidence angle and the coupling among elements. Moreover, our approach also considers the effects of the interaction of the periodic surface with the RIS ground plane. The model allows also to consider the behavior of the RIS as a function of the incidence angle. We will show the accuracy of the model by considering an example of RIS which connects a receiver to a transmitter. 

The paper is organized as follows. In Section \ref{sec_Model} the addressed problem is discussed and motivated. Section \ref{sec_Metasurface_model} presents the homogenization approach for analysing a metasurface and its equivalent transmission line circuit. Section \ref{sec_RIS_model} describes the transmission line mode of the RIS. The accuracy of the RIS model is verified by comparing the reflection coefficients with electromagnetic simulations in Section \ref{sec_results}. The model is employed in Section \ref{sec_link} where a practical example of RIS-assisted communication is analysed. Finally conclusions are drawn in Section \ref{sec_conclusion}.

\section{Addressed Problem} 
\label{sec_Model} 

The focus of this work is to provide an accurate and simple approach for determining the possible values of amplitude and phase of the RIS based on the physical constraints of the hardware implementation. The physically based model takes into account the properties of RIS which are the reconfigurability, the oblique incidence properties (spatial dispersion), the mutual coupling among closely spaced cells and reflection losses. Moreover, the model can guarantee a direct link between the reflection phases and the physical layout of the RIS. 

In order to focus a significant amount of the power towards the user, the RIS dimensions might be considerably large because the collected power is related to the physical area of the reflecting surface. Consequently, it is not uncommon that both the transmitter and the received can be located in the near field of the surface \cite{ellingson2019path,bjornson2020power}. Therefore it is of vital importance that the model of the surface is able to consider the specific incident and reflected angle for each unit cell.  

\subsection{Reconfiguration methods}
\begin{figure}
	\centering
	\includegraphics[width=1\linewidth]{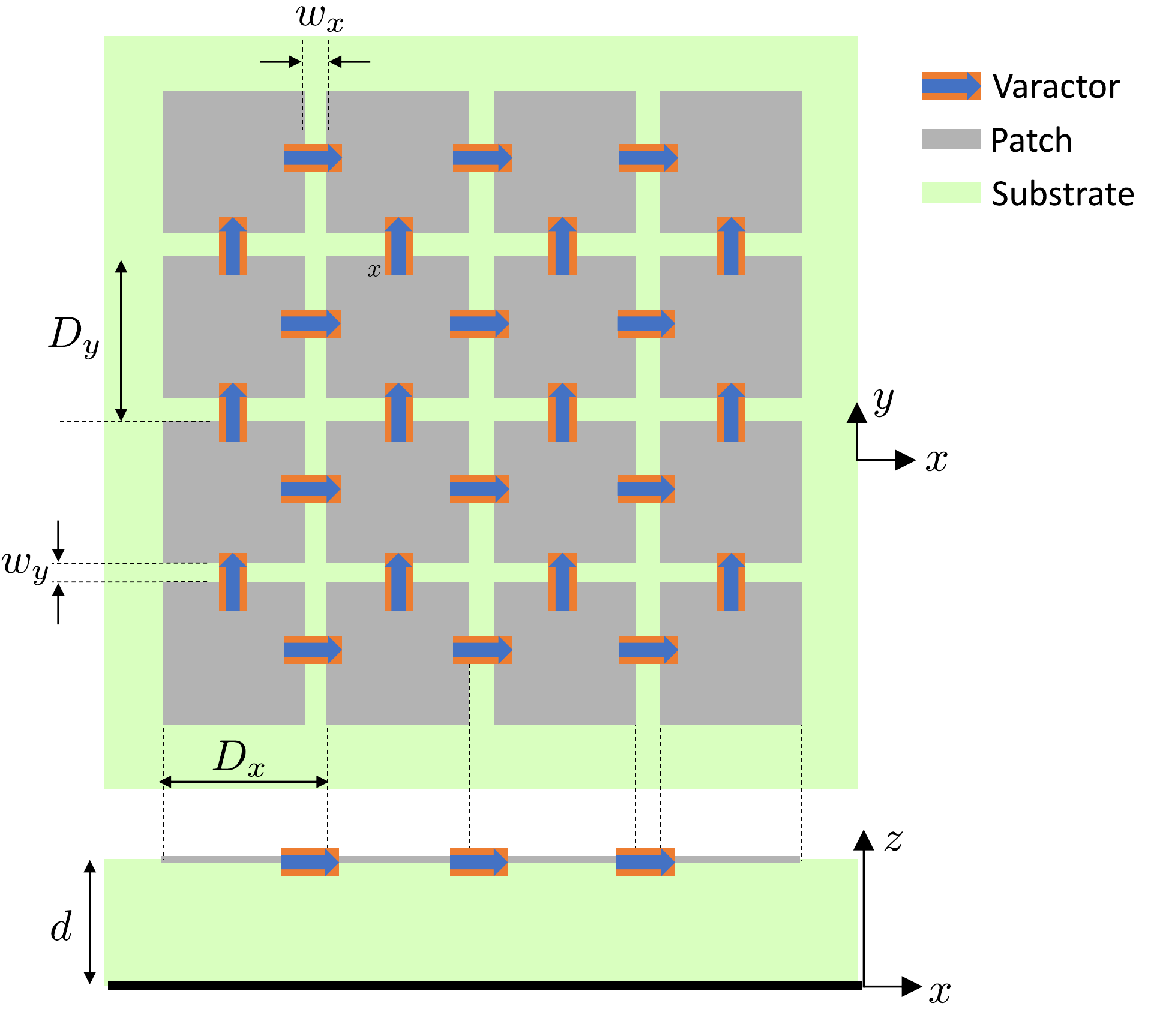}
	\caption{Geometry of the Reconfigurable Intelligent Surface. $D_x$,  $D_y$ are the periodicity of the lattice and $ w_x $ and $ w_y $ are the gaps between the squares patches along \textit{x} and \textit{y} planar directions. Varactor diodes connect adjacent cells in both planar directions. }
	\label{fig_AIS_geometry}
\end{figure}

RIS are synthesized through thin resonant cavities that fully reflect incident waves with an arbitrary reflection phase. The the full reconfigurability of the RIS can be achieved in various way but the simplest approach is based on the adoption of varactor diodes within each unit cell as shown in Fig.~\ref{fig_AIS}. The physical aspect of the structure with an periodic surface composed by square metal patches loaded with varactor diodes is shown in Fig.~\ref{fig_AIS_geometry}. When an Artificial Impedance Surface (AIS) is loaded with active components, the reflection properties of each unit cell of the surface can be controlled as a function of a direct current (DC) voltage. By adequately changing the polarization state of the varactors, it is possible to perform the beamforming so that each user can be reached with the maximum available power.  
To this aim, in the microwave regime, varactor \cite{costa2008active} or PIN diodes \cite{genovesi2012frequency}, whose EM characteristics (say, capacitances) can be dramatically tuned through varying applied voltages, can be incorporated within the surface. By independently controlling the voltages of each unit cell, phase/amplitude profiles can be accurately synthesized in order to shaping and controlling the reflected fields. Examples of varactor-controlled electrically steerable reflector in the microwave regime can be found since the early 2000s	\cite{sievenpiper2003two, sievenpiper2005forward, liswitchable}. Alternative approaches to tune EM properties of metadevices rely on the free carrier doping in conductive materials with electrical gating and photoexcitation methods. Conventional semiconductors (e.g., GaAs, Silicon, and germanium), atomically thin 2D materials (e.g. graphene, MoS2), and transparent conducting oxides or nitrides such as Indium Thin Oxide (ITO), Aluminum-doped Zinc Oxide (AZO) are widely used in the spectral range covering THz and up to the visible regime \cite{he2019tunable, tsilipakos2020toward}. Mechanical tuning offers another effective way to switch the EM properties of metadevices by reconfiguring the shape and surrounding environment of meta-atoms by the means of MEMS/NEMS, elastic subtract, or microfluidics \cite{he2019tunable, tsilipakos2020toward, wang2020theory}.

\subsection{Oblique Incidence}
\begin{figure}
	\centering
	\def\svgscale{1.3}
	\subfloat[]{%
		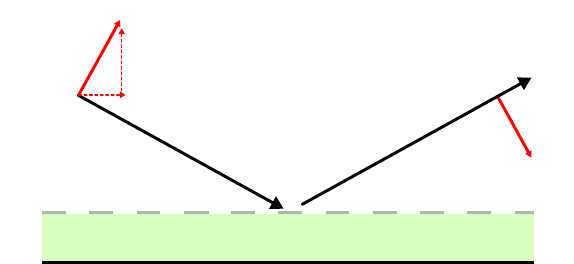}
\hfill
	\def\svgscale{1.3}
	\subfloat[]{
		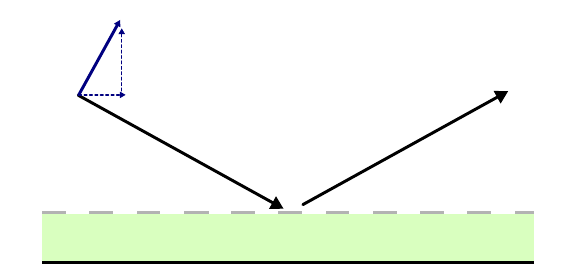}
	\caption{Oblique incidence on a RIS. (a) Parallel polarization (or Transverse Magnetic  - TM) (b) Perpendicular polarization (or Transverse Electric - TE). $\vec{E}^i$ and $\vec{H}^i$ are the impinging electric and magnetic fields respectively whereas $\vec{E}^s$ and $\vec{H}^s$ are the scattered electric and magnetic fields. $\theta^i$ and $\theta^s$ are the incident and scattered angle of the impinging wave.}
	\label{fig_polarization}
\end{figure}

One important aspect of these structures is their working mechanism which is based on the interference between the periodic metasurface and the ground plane. Consequently, the reflection properties of an electromagnetic wave striking upon a RIS, strongly depend on the incidence angle. In electromagnetic language, it is said that the device is spatially dispersive. This aspect is completely overlooked in most of the papers published on this subject. For this reason, a physics based but still simple model is extremely important for the correct design of the RIS. Two meaningful situations can be distinguished when an electromagnetic wave impinges on the surface: parallel polarization (TM - Transverse Magnetic) and perpendicular polarizations (TE - Transverse Electric). At normal incidence, for a symmetric surface topology, the behaviour of the surface for parallel and perpendicular polarization coincides. However, for off-normal angles the behaviour of the RIS for TE and TM polarization should be analyzed separately. In Fig.~\ref{fig_polarization} the parallel and perpendicular polarization incidence are shown. In the parallel polarization, the electric field lays in the plane of incidence whereas the magnetic field is orthogonal to the same plane. On the contrary, for the parallel polarization case, the magnetic field lays in the plane of incidence whereas the electric field is orthogonal to the same plane.

\subsection{Mutual Coupling}
Another key factor in the analysis of RIS is the effect of the mutual coupling. The electromagnetic analysis of RIS is based on the local periodicity approach  \cite{pozar1997design, nayeri2013radiation, Borgi_reflectarray}. It is indeed assumed that the size of the characteristics of the elements varies gradually along the surface and this justify the approximation of studying a single unit cell response with an infinite set of identical elements. Once a parametric simulation of the elements with different sizes is performed, a database of phases is created and thus a correspondence between geometrical parameters and phase response of a particular pixel of the periodic surface can be created. The analysis of the phase response of the periodic surface for a certain element configuration has a twofold advantage. The first advantage is that the periodic surface can be analysed by using the Floquet theorem which consider the single unit cell placed in an infinite array of elements all identical to each other. This approach, is much more efficient (in terms of computational time required for the EM simulation) than studying an isolated unit cell. The second advantage is that the mutual coupling between neighboring unit cells is automatically included in the calculations because the unit cell is analysed in an infinite array and not in a standalone configuration. For this reason, we present in the following section an approach for the modelling of a periodic Impedance surface configuration. More details regarding the Floquet theorem are provided in Appendix-\ref{Appendix-a}

\subsection{Reflection Losses}
A physical surface with a phase-controlled reflection coefficient is strongly affected by losses. The losses in the surface depend on multiple factors: dielectric losses due to the use of real materials (glass-reinforced epoxy laminate material such as FR4 are cost-effective but they also exhibit	 highly losses), ohmic losses due to the current flowing on imperfect conductors and losses in the active components. It is important to underline that an efficient strategy to limit the effect of losses consists in keeping small the periodicity of the lattice \cite{costa2012closed, ethier2012loss, pozar2007wideband}. Moreover, the adoption of a pattern with small period is also the best strategy for achieving wideband RIS \cite{pozar2007wideband, costa2009bandwidth}. The usual practice of designing reflectarrays with $\lambda/2$ spaced elements is not the best configuration for the bandwidth maximization or for loss minimization. However, reducing the period determines the increase of the number of active components on the periodic surface required to achieve reconfigurability. Consequently, a suitable compromise between these two requirements should be found. 

\section{EM Model of Metasurface} 
\label{sec_Metasurface_model}

The periodic surface employed for the design of RIS is essentially a two–dimensional array comprising disconnected metal obstacles. If the periodicity of the periodic surface is much larger than the operating wavelength $\lambda$, the periodic surface can be efficiently analyzed by using the homogenized theory \cite{lukkonen_simple_model, yang2019surface}. In order to avoid the propagation of high-order Floquet harmonics due to the lattice of the periodic surface, the operating frequency ($f_0$) should satisfy the following condition:

\begin{equation} \label{eq_HO_modes} 
	f_0 < \frac{c}{D (\sqrt{\epsilon_r}+\sin{\theta^i})}
\end{equation}

where $D$ represents the periodicity of the periodic surface, $\epsilon_{eff}$ is the effective dielectric permittivity of the surrounding medium \cite{costa2021simple}, and $\theta^i$ is the incident angle of the impinging electromagnetic wave. If (\ref{eq_HO_modes}) is satisfied, higher order Floquet modes (grating lobes) are in cut-off and only the first order response of the surface can be considered. Consequently, according to the averaged theory, the periodic surface can be studied as an homogeneous interface characterized by a finite surface current. Therefore, the periodic surface can be described as a uniform sheet characterized by a frequency dependent surface impedance. The behaviour of the surface impedance as a function of frequency depends on the shape of the element. A patch type surface for instance is characterized by a pure capacitive behaviour whereas a loop type or a cross type element is characterized by a resonant behaviour which can be fitted through a LC series circuit \cite{costa2014overview}.  In general, if the surface is characterized by a bulk profile with the presence of electric and magnetic dipole moments, a more complete characterization of the metasurface can be accomplished through the  generalized sheet-transition condition \cite{holloway2012overview}. In this way, the metasurface is modelled as an electric and a magnetic impedance applied across an infinitely thin equivalent surface. Since in our case the surface is infinitesimally thin, only the electric impedance differs from zero. As a consequence, the electric field is continuous at the interface and the magnetic field jumps. The derivation of the reflection coefficient of a freestanding homogenized surface for an obliquely incident plane wave is detailed in Appendix~\ref{Appendix-b}. It is shown that the derivation obtained by applying the impedance boundary condition in eq.~(\ref{eq_GSTC}) is equivalent to the transmission line approach. The latter is employed in the following section to derive the equivalent circuit model of the RIS.

\section{EM Model of RIS} 
\label{sec_RIS_model} 

The not reprogrammable version of the surface is also known as Artificial Impedance Surface (AIS) \cite{costa2009bandwidth, lukkonen_simple_model, maci2005pole}. The AIS comprises a periodic surface printed on the top of a grounded dielectric slab.  
  
\begin{figure}
	\centering
	\includegraphics[width=1\linewidth]{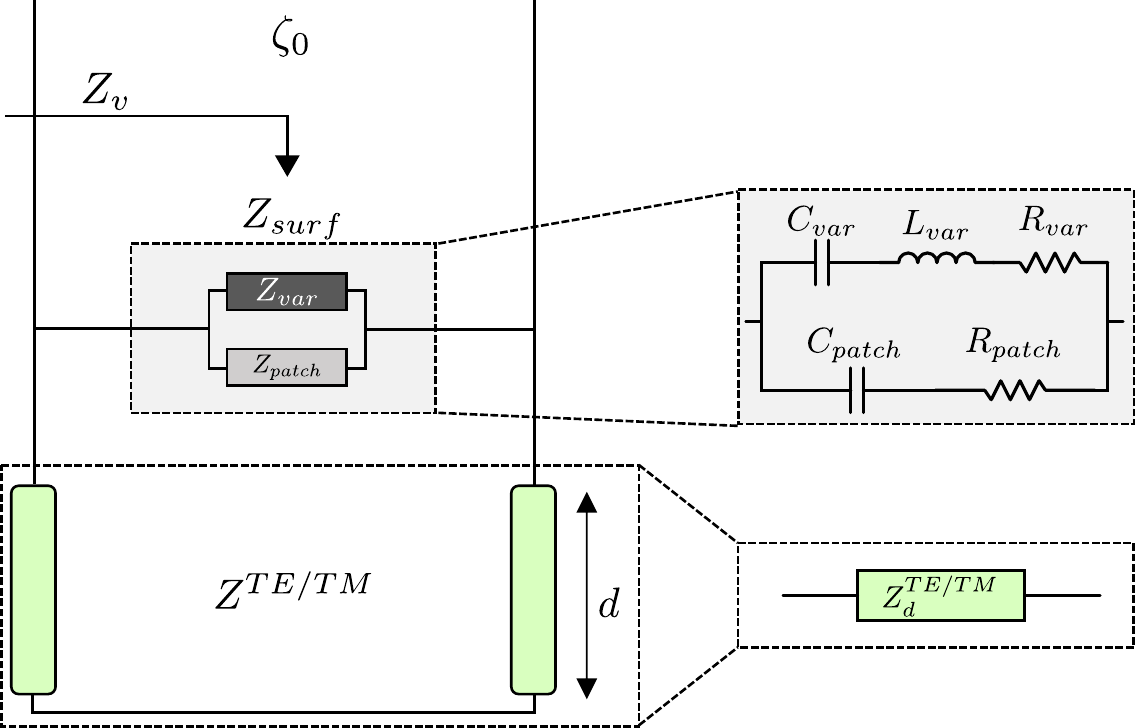}
	\caption{Equivalent circuit model of a RIS. $Z_{v}$ is the input impedance of the transmission line model, $Z_{surf}$ represents the total surface impedance of the periodic surface loaded with varactors. $Z_{patch}$ is the surface impedance of the patch array, $Z_{var}$ represents the lumped impedance of the varactor diode), $d$ is the length of the transmission line section which represents the substrate, $\zeta_0$ is the characteristic impedance of free space and $Z^{TE/TM}$ is the characteristic impedance of the substrate for TE or TM incidence.}
	\label{fig_eq_circuit}
\end{figure}

The conventional transmission line model of an artificial impedance surface consists in a parallel connection between the surface impedance of the metasurface and inductive impedance of a short transmission line closed on a short circuit \cite{lukkonen_simple_model, costa2012circuit}. As shown in Fig.~\ref{fig_eq_circuit}, the input impedance of the AIS structure ($Z_v$) is equal to the parallel connection between impedance of the periodic surface $Z_{surf}$ and the impedance of the grounded dielectric slab $Z_d^{TE/TM}$: 

\begin{equation} \label{eq_Zv} 
	Z_{v}=\frac{Z_{surf} \; Z_d^{TE/TM}}{Z_{surf}+Z_d^{TE/TM}}
\end{equation}

According to the equivalent circuit model in Fig.~\ref{fig_eq_circuit}, $Z_{surf}$ is computed as the parallel connection of the unloaded surface impedance of the patch array, $Z_{patch}$, and the lumped impedance of the varactor diode, $Z_{var}$. Once the input impedance of the structure is computed, the complex reflection coefficient can be calculated as follows \cite{lukkonen_simple_model}: 

\begin{equation} \label{eq_Zv} 
	\Gamma=\frac{Z_{v} - \zeta_0}{Z_{v}+\zeta_0}
\end{equation}

Two meaningful cases should be distinguished when the reflection coefficient is calculated for oblique incidence: parallel polarization (TM - Transverse Magnetic) in which the electric field lies on the plane of incidence and the perpendicular polarization (TE - Transverse Electric) in which the electric field is perpendicular with respect to the plane of incidence. The two oblique incidence cases are represented in Fig.~\ref{fig_polarization}. 

The thin grounded dielectric slab behaves as an inductor. Consequently, its impedance can be computed analytically as follows \cite{lukkonen_simple_model,costa2012closed}:

\begin{equation} \label{eq_Zd} 
	Z_{d}^{TE/TM}=jZ^{TE/TM}\tan{(k_{z}d)};
\end{equation}

where $Z^{TE/TM}$ represents the characteristic impedance of the substrate along the equivalent transmission line towards $z$ direction:

\begin{equation}
	Z^{TE}=\frac{\omega \mu}{k_{z}} \;\; ; \;\; Z^{TM}=\frac{k_{z}}{\omega \varepsilon_0 \varepsilon_r}
\end{equation}

$k_{z}=k_0 \sqrt{\varepsilon_r - \sin^2\theta^i}$ is the normal propagation constant within the substrate, $k_0$ is the propagation constant of the incident plane wave and $\theta^i$ is the incidence angle. At normal incidence the TE and TM case equals to each other and the impedance of the medium is simply $\zeta_0/ \sqrt{\varepsilon_r}$, where $\zeta_0$ is the free space impedance ($\zeta_0 = \sqrt[]{\mu_0 / \varepsilon_0} $), where $\varepsilon_0$ and $\mu_0$ are the free space dielectric permittivity and magnetic permeability, respectively. However, at oblique incidence two cases must be analysed since the behaviour of the surface change drastically depending on the polarization of the field \cite{lukkonen_simple_model, costa2012circuit}. 
The impedance of a periodic surface can be represented through a lumped equivalent circuit if the periodicity of the lattice is lower than one wavelength. The specific geometry of the unit cell of the periodic surface determines the impedance properties of the surface and its circuit representation. Usually, the capacitance and the inductance are determined by the shape of the periodic element while the resistance is influenced by the losses of the metal of the printed structure. For instance, the impedance of an array of patches ($Z_{patch}$) can be represented by a capacitance only up to a certain frequency where the granting lobe onset:

\begin{equation} \label{eq_ZFSS} 
	Z_{patch}^{TE/TM}=R_{patch}+\frac{1}{j \omega C_{patch}^{TE/TM}}
\end{equation}
	
On the contrary, more complicated geometries have also a considerable inductive part in the surface impedance \cite{costa2014overview}. 

It is important to highlight that the impedance of a periodic structure strongly depends on the incidence angle ($\theta^i$) of the EM wave impinging on the AIS. For the case of periodic patches arrays, the capacitance for the TE and the TM polarization can be expressed as \cite{lukkonen_simple_model, costa2011design}:

\begin{equation} \label{eq_CTE} 
	\resizebox{.91	\hsize}{!}{$C^{TE}_{patch}= \frac{2D_y\varepsilon_0\varepsilon_{eff}}{\pi}\ln{\left( \frac{1}{\sin\left(\frac{\pi w_y}{2D_y}\right)}\right) } \left(1-\frac{k_0}{k_{eff}}\frac{\sin^2\theta^t}{2}\right)$}
\end{equation}

\begin{equation} \label{eq_CTM} 
	C^{TM}_{patch}= \frac{2D_x\varepsilon_0\varepsilon_{eff}}{\pi}\ln{\left(\frac{1}{\sin\left(\frac{\pi w_x}{2D_x}\right)}\right)}
\end{equation}

where $\varepsilon_{eff}=(1+\varepsilon_r)/2$ is the effective permittivity of the uniform medium surrounding the periodic surface, $k_{eff}$ is the wave number of the incident wave vector
in the effective host medium. $D_x$,  $D_y$ are the periodicity of the lattice and $ w_x $ and $ w_y $ are the gaps between the squares patches along \textit{x} and \textit{y} planar directions.
The averaged formulas can be employed for computing the reflection coefficient both for normal and oblique incidence.  The presented patches arrays model which is analyzed for the case of azimuthal incidence angle equal to zero, can be extended to a generic angle due to the geometrical symmetry of the unit cell.
The aforementioned averaged expressions of the capacitance refer to cases in which the dielectric thickness is sufficiently thick so that high-order Floquet harmonics can be neglected \cite{costa2012circuit}. In order to take into account the effect of higher-order Floquet harmonics the capacitance of the periodic surface can be corrected as \cite{costa2012circuit}:

\begin{equation} \label{eq_C_refined} 
	C^{TE/TM}_{patch}= C^{TE/TM}_{patch} - C_{patch-ground}
\end{equation}

where $C_{patch-ground}$ is a correction terms which takes into account the capacitance between the patch and the ground plane:

\begin{equation} \label{eq_cap_correction} 
	C_{patch-ground}= \frac{2D \varepsilon_0\varepsilon_{r}}{\pi} \ln{(1- e^{-4\pi d/D})}
\end{equation}

when $D_x = D_y = D$. If the substrate thickness $d$ becomes much smaller than the cell periodicity $D$, the influence of higher‐order (evanescent) Floquet modes reflected by the ground plane starts to play an important role and, when the ratio $d/D$ approaches to zero, the correction term tends to infinite as well as the capacitance of the patch array. If $d$ is small compared to the wavelength but large compared to $D$, the interaction between the grid and the metal plane is negligible. 

The analysis of the RIS	 surface is performed by adding an additional lumped capacitor, in parallel with the capacitance created by the patch array \cite{costa2011design}. The effect of the diode is taken into account by placing the lumped element impedance in parallel to the impedance of the periodic surface. As is well known, a varactor diode can be represented as a series of a resistor, inductor and a capacitor: 

\begin{equation} \label{eq_varactor} 
	Z_{var}=R_{var} + j \omega L_{var} + \frac{1}{j \omega C_{var}};
\end{equation}

where $C_{var}$ represents the variable capacitance of the diode. The series inductance $L_{var}$ depends on the size of the lumped component and must be included in the varactor model \cite{note2010varactor} to take into account the self resonance of the component. The losses of the varactor are taken into account by the series resistor $R_{var}$. The bandwidth of the structure is proportional to equivalent inductance of the grounded substrate. Consequently, the thicker the substrate, the larger the operating bandwidth and therefore the smother the phase slope as a function of frequency. Concerning the element shape, the patch with a vary small edge gap represents the optimal solution to maximize the operating bandwidth of the AIS \cite{costa2009bandwidth}.

Differently from \cite{tang2020mimo} or \cite{najafi2020physics} where a physical model for the RIS impedance (named $Z_{n,m}$)  \cite{tang2020mimo} is not available, the proposed model allows to consider the geometrical and electrical characteristics of the RIS in the computation of the reflection coefficient of the unit cell. The proposed approach clarifies the connection between  and the hardware implementation of the RIS including varactor diodes, cell topology, dielectric substrates and the effect of oblique incidence.

\section{Numerical results} 
\label{sec_results} 

\begin{figure}
	\centering
	\subfloat[]{%
		\includegraphics[height=0.63\linewidth,trim={33cm 0 0 0},clip]{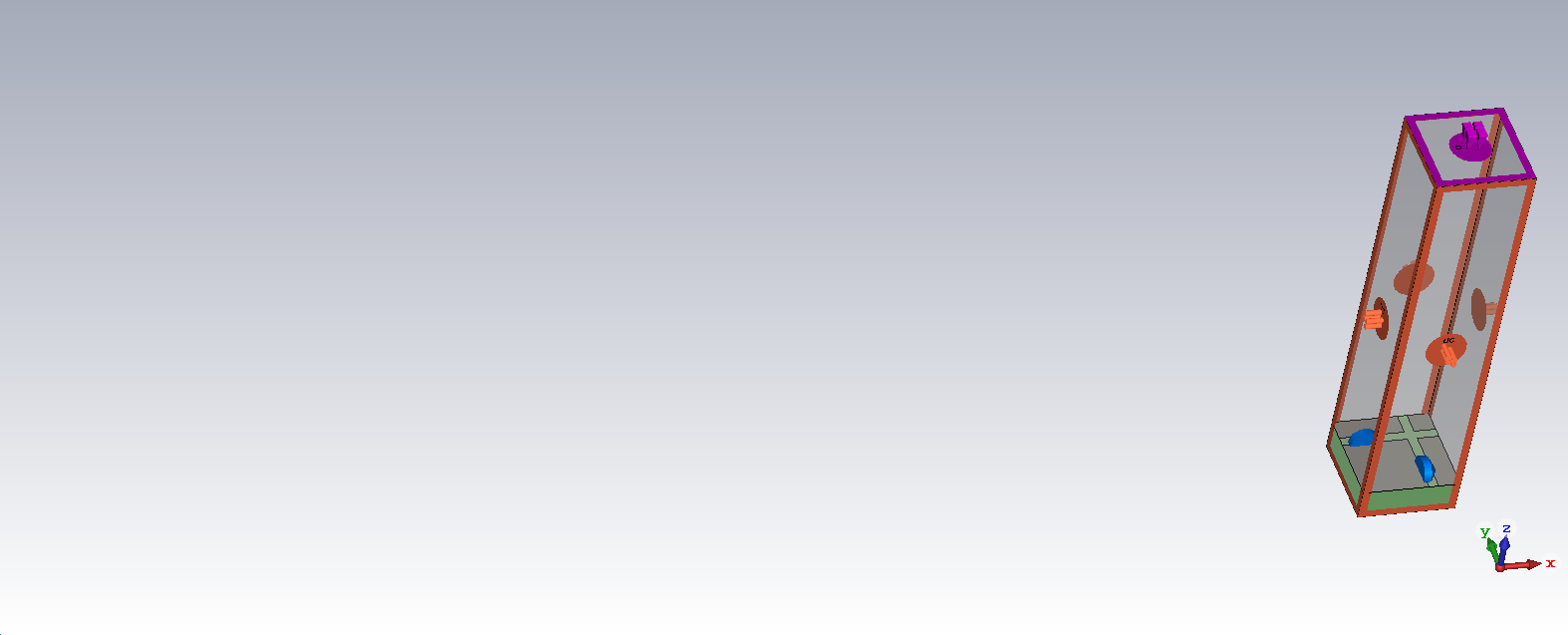}}
	\hfill
	\subfloat[]{%
		\includegraphics[height=0.63\linewidth,trim={28cm 0 0 0},clip]{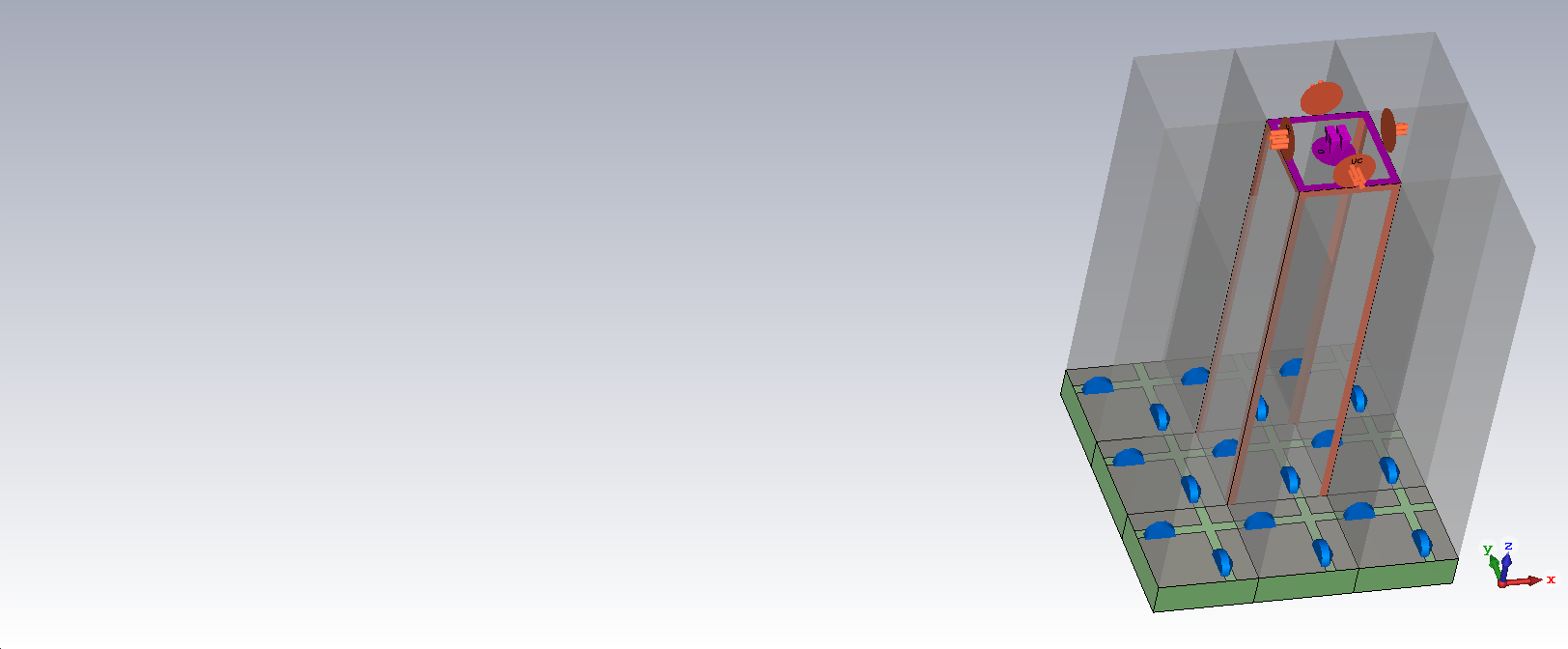}}
	\hfill
	\caption{3D layout of the Varactor loaded RIS simulated with CST Microwave Studio. The simulation relies on the Floquet theorem which considers the unit cell placed in an infinite periodic array: (a) analysed unit cell, (b) equivalent infinite array.}
	\label{fig_CST_model}
\end{figure}

\begin{figure}
	\centering
	\subfloat[][\scalebox{0.9}{\textit{TE-amplitude}}]{%
		\includegraphics[width=4cm,keepaspectratio]{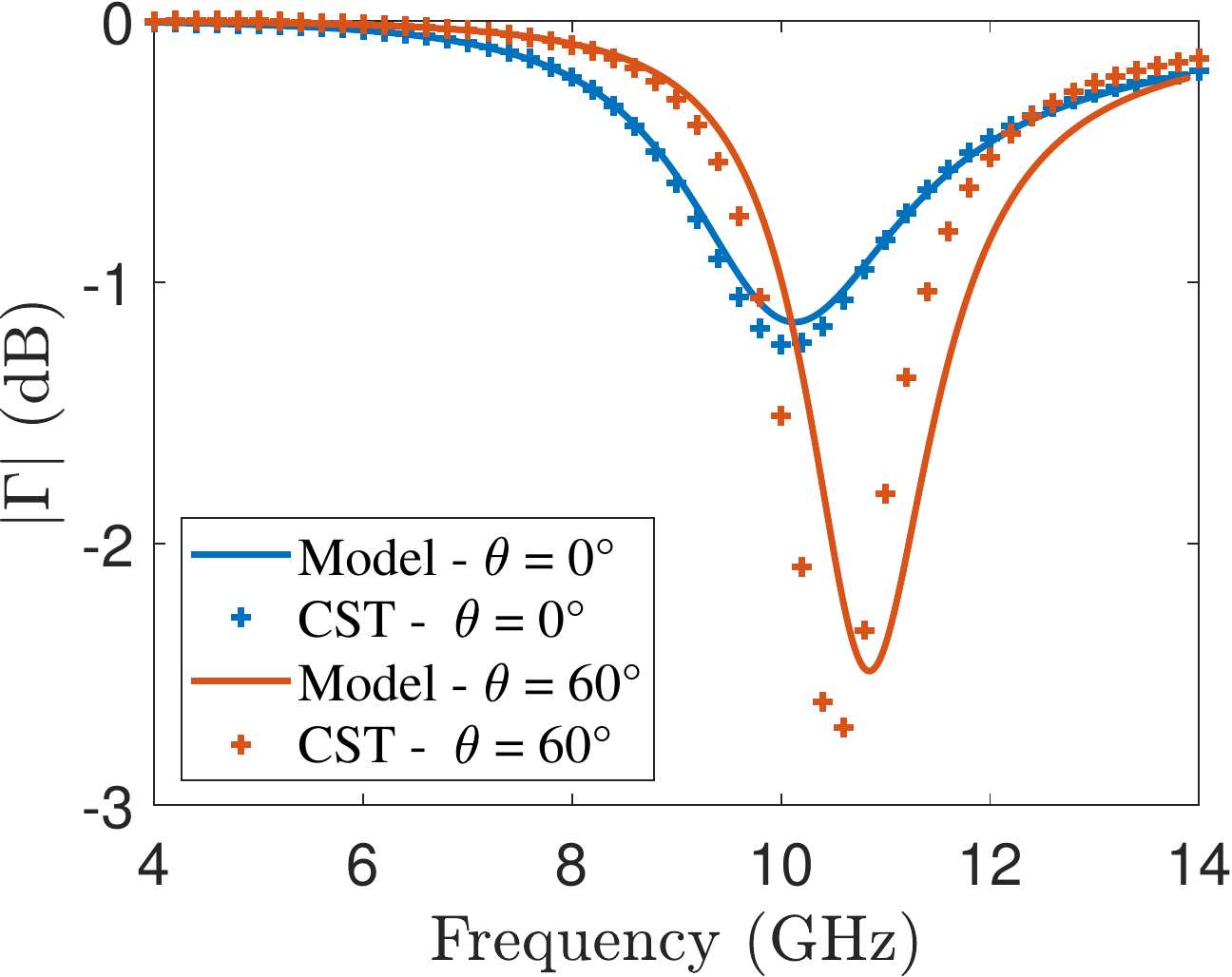}}
	\label{}\hfill       
	\subfloat[][\textit{\scalebox{0.9}{\textit{TM-amplitude}}}]{%
		\includegraphics[width=4cm,keepaspectratio]{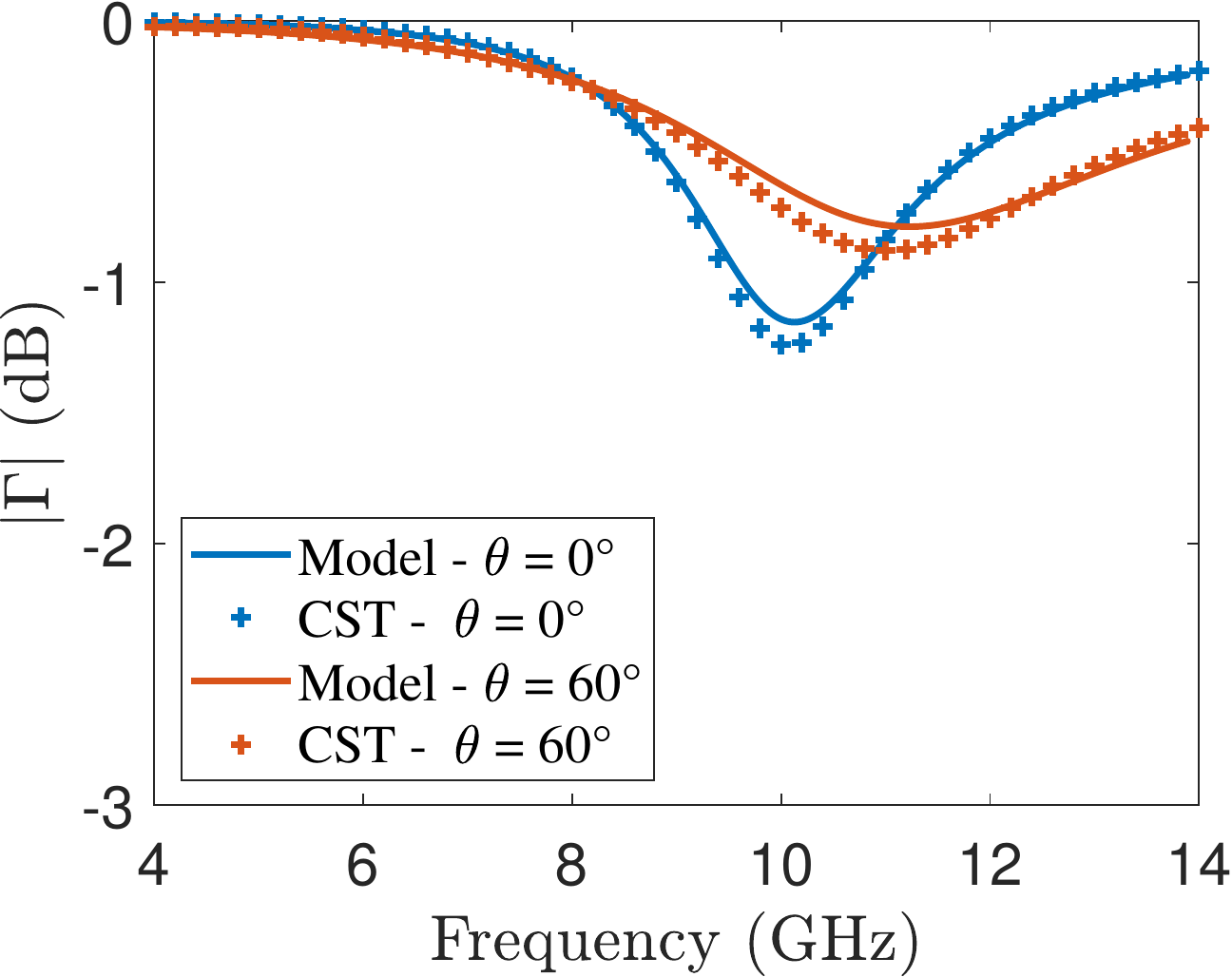}}
	\label{}\hfill       
	\subfloat[][\textit{\scalebox{0.9}{\textit{TE-phase}}}]{%
		\includegraphics[width=4cm,keepaspectratio]{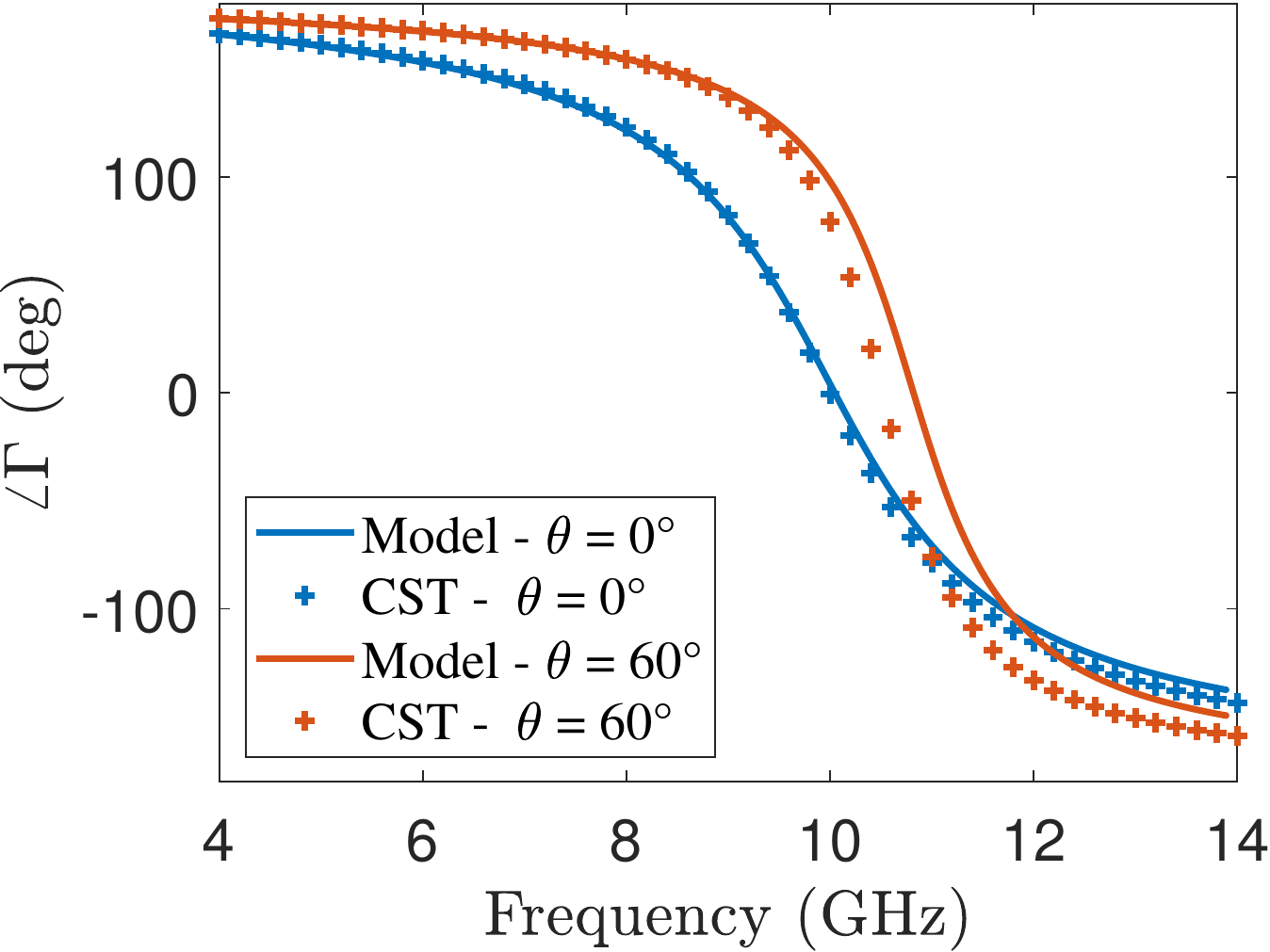}}
	\label{}\hfill        
	\subfloat[][\textit{\scalebox{0.9}{\textit{TM-phase}}}]{%
		\includegraphics[width=4cm,keepaspectratio]{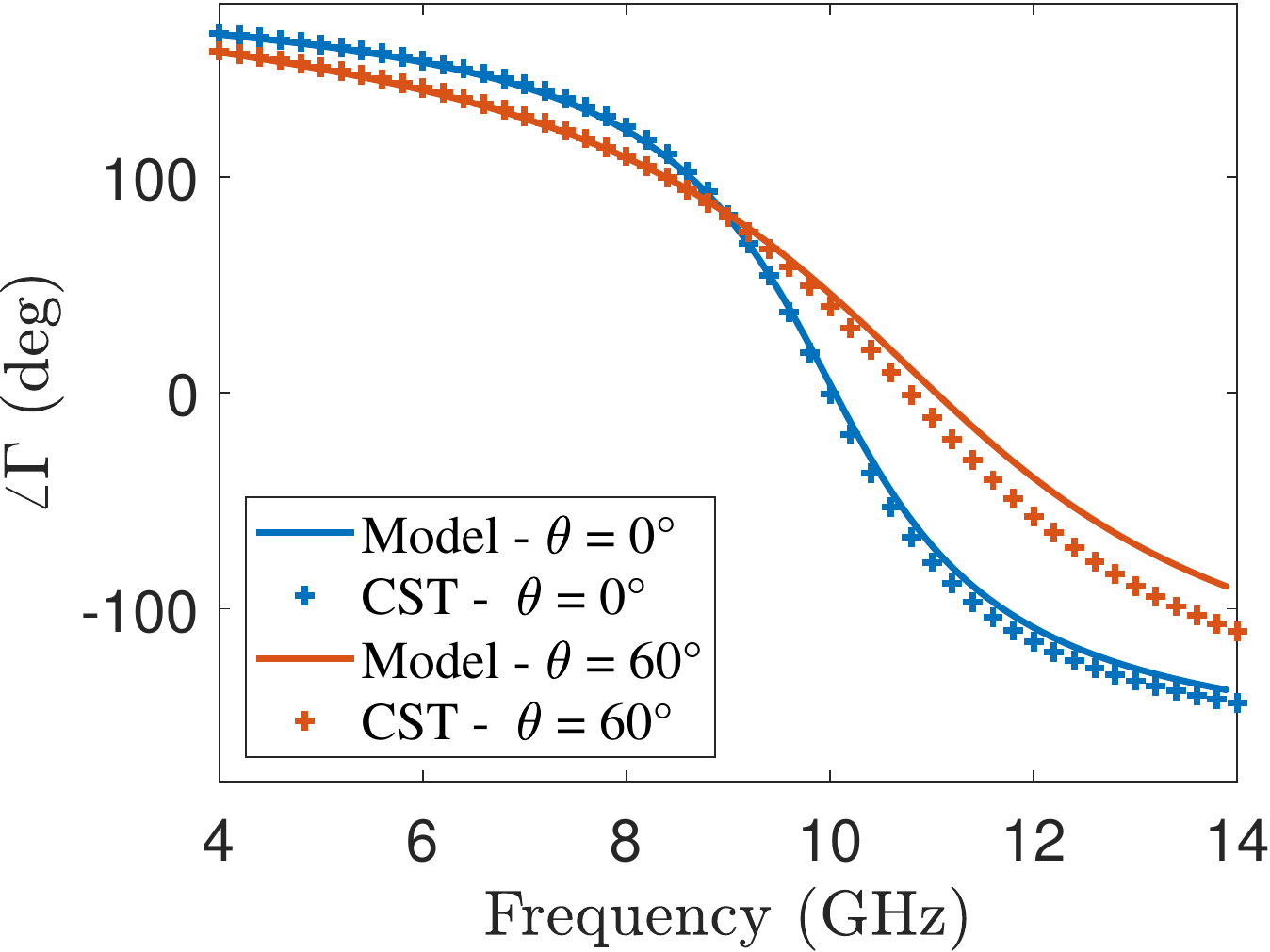}}
	\label{}\hfill        
	\caption{(a) TE and (b) TM reflection phase for normal ($\theta = 0^\circ$) and oblique incidence ($\theta = 60^\circ$). The equivalent TL model is compared with the results obtained with CST Microwave Studio. The geometrical and electrical parameters used for the AIS are the following: $D_x = D_y = D = 5$ mm, $w_x = 0.5$ mm,	$\varepsilon_r = 4.4-j0.088$, $d = 1.2$ mm.}
	\label{fig_reflection_phase_AIS} 
\end{figure}

In order to assess the accuracy of the proposed circuit model, an example of reconfigurable surface is analysed. The selected parameters are the following: $D=5$ mm , $d=1.2$ mm, $w=0.5$ mm and $C_{var}=[0.1-0.5]$ pF. The reflection phase of the AIS without the varactor diode is initially analysed at normal and oblique incidence both by using the proposed model and by using a commercial electromagnetic simulator, that is CST Microwave Studio. CST simulations for Reconfigurable Impedance Surfaces (RIS) has been verified several times in the electromagnetic community as for instance in \cite{sievenpiper2003two, costa2008active, luukkonen2010experimental}. The layout of the unit cell analysed with CST is shown in  Fig.~\ref{fig_CST_model}. The amplitude and phase reflection coefficient of the surface are reported in Fig.~\ref{fig_reflection_phase_AIS} for normal and oblique incidence. As evident, the agreement between the full-wave results and the proposed model is satisfactory for both normal and oblique incidence with only a small frequency shift between CST simulations and the model. However, it has to be pointed out that the shift diminishes as the discretization mesh is refined in the electromagnetic solver. The main hurdle in computing the Radar Cross Section (RCS) of these finite structures hosting several resonators placed close to a ground plane is that the resonators  and the metallic surface act as a Fabry-Perot interference device with multiple reflection contributions involved. In order to accurately capture these phenomena, a very fine local mesh is required to achieve good results. The next step is to introduce the varactors which allows the active control of the properties of the AIS. In this case, the surface can be classified as Intelligent Reflective Surface since the reflection phase properties can be controlled an each unit cell by though a DC voltage. Each unit cell should be controlled independently in order to guarantee a complete reconfiguration. An architecture for achieving the full control of the elements in the two planar direction was presented in \cite{sievenpiper2003two}.

\definecolor{mycolor}{RGB}{255,204,170}

\begin{figure}
	\centering
	\subfloat[][\scalebox{0.9}{\textit{amplitude}}]{%
		\includegraphics[width=4cm,keepaspectratio]{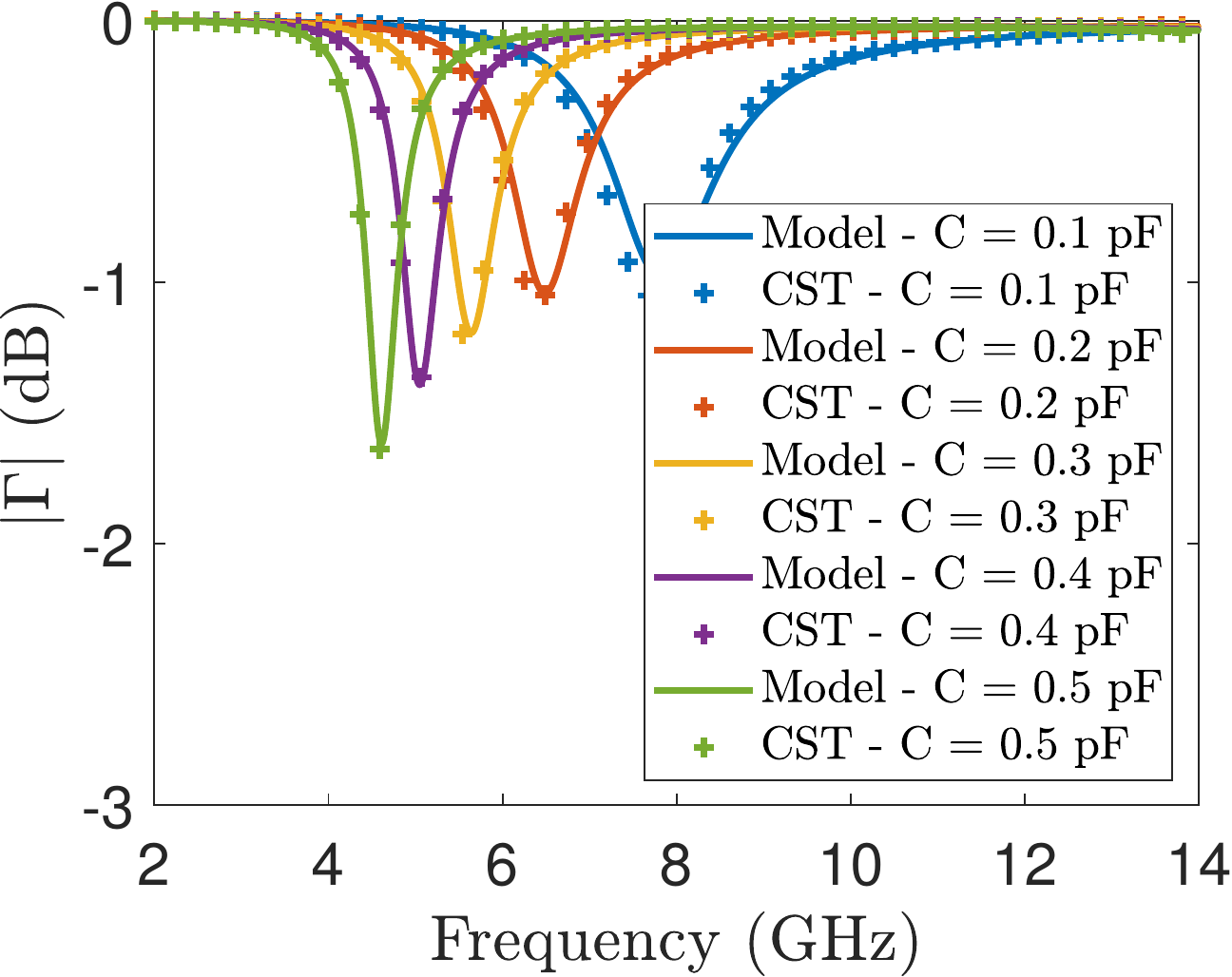}}
	\label{}\hfill        
	\subfloat[][\textit{\scalebox{0.9}{\textit{phase}}}]{%
		\includegraphics[width=4cm,keepaspectratio]{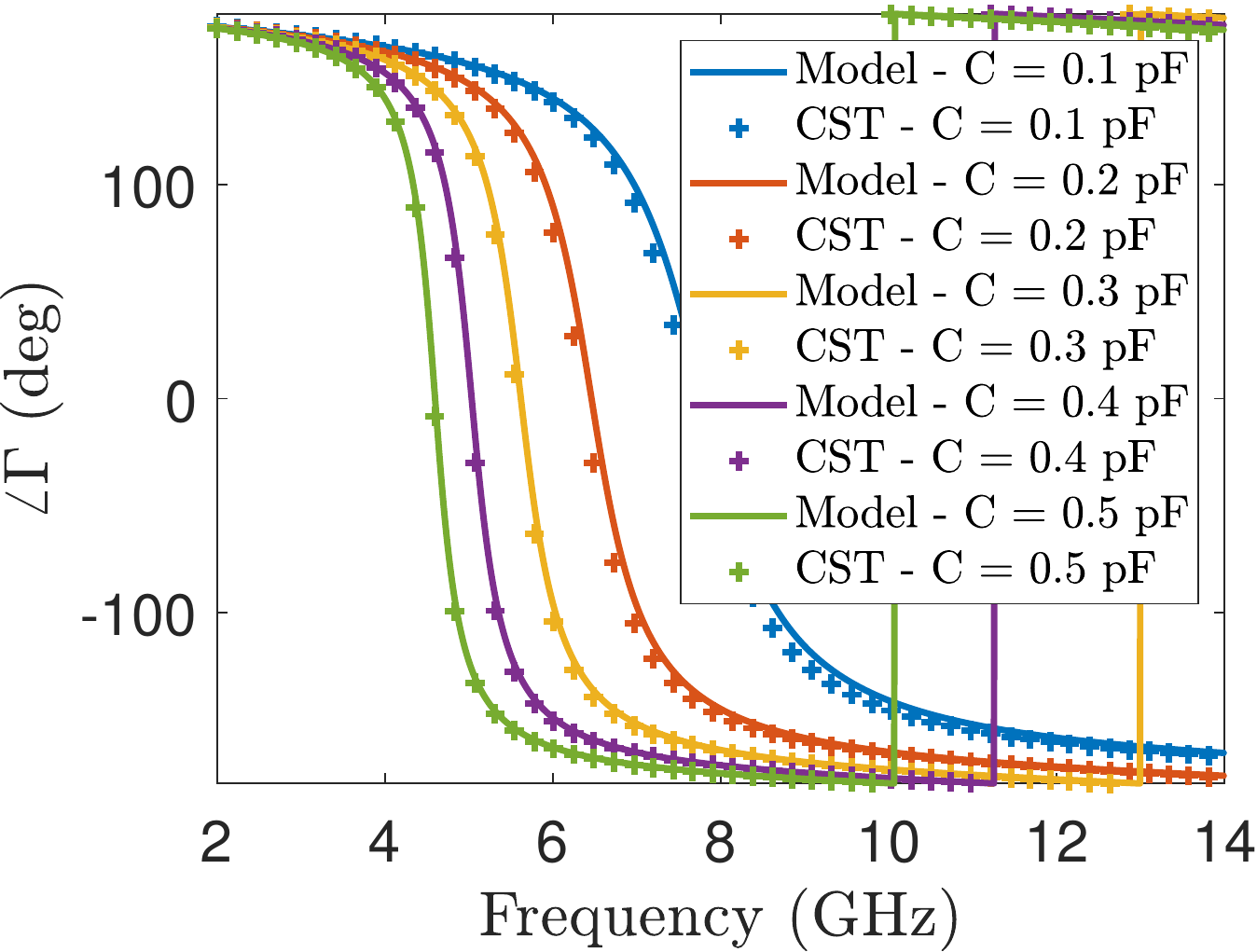}}
	\label{}\hfill        
	\caption{(a) Amplitude and (b) phase of the reflection coefficient at normal incidence for a RIS composed by a varactor loaded AIS. The results of the model are compared with CST Microwave Studio. The geometrical and electrical parameters used for the RIS are the following: $D_x = D_y = D = 5$ mm, $w_x = 0.5$ mm,
		$\varepsilon_r = 4.4-j0.088$, $d = 1.2$ mm, $R_{var} = 0.5 \Omega $.}
	\label{fig_reflection_phase_RIS} 
\end{figure}

\begin{figure}
	\centering
	\subfloat[][\scalebox{1}{\textit{Real Part}}]{%
		\includegraphics[height=3.3cm,keepaspectratio]{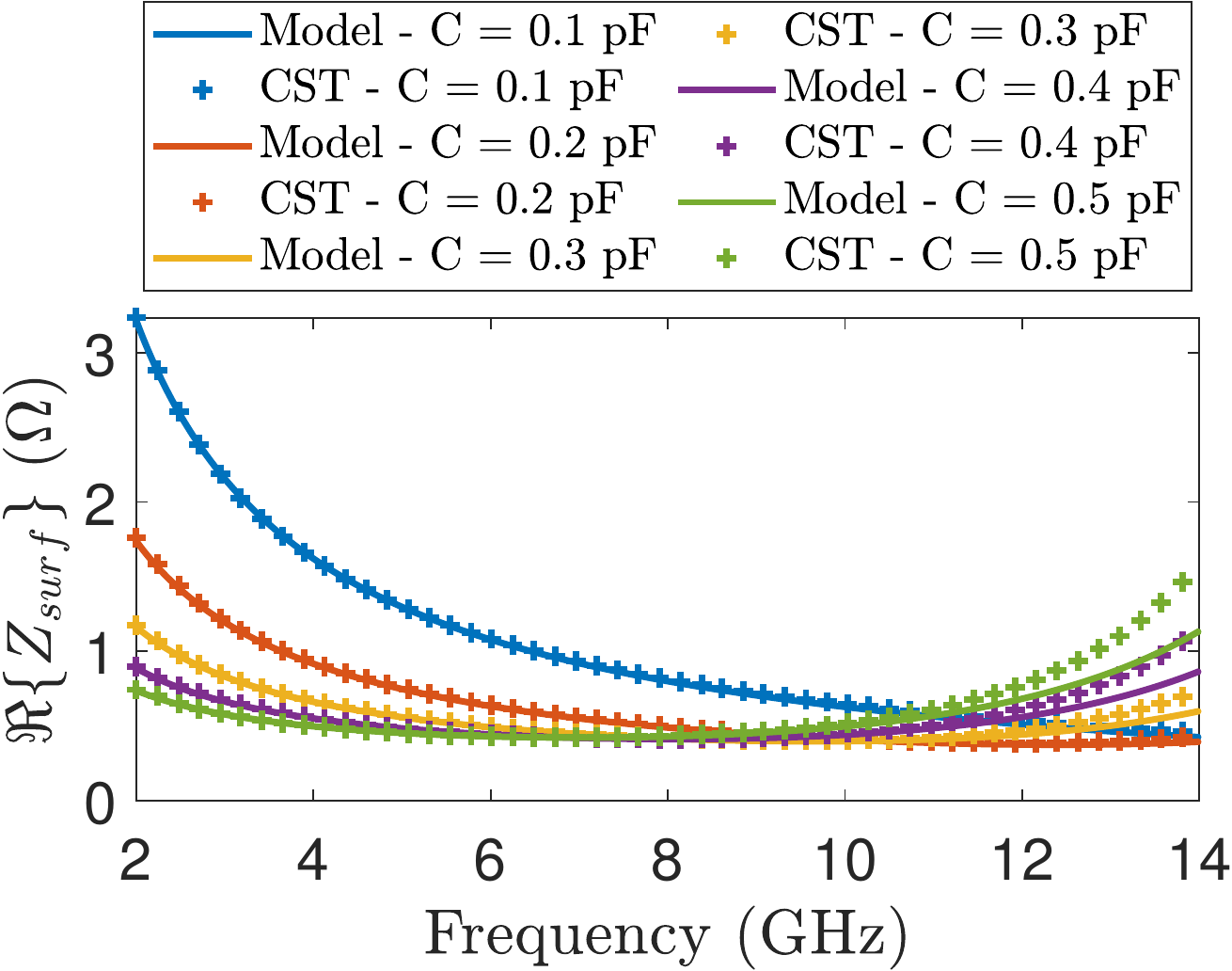}}
	\label{}\hfill        
	\subfloat[][\textit{\scalebox{1}{\textit{Imaginary part}}}]{%
		\includegraphics[height=3.3cm,keepaspectratio]{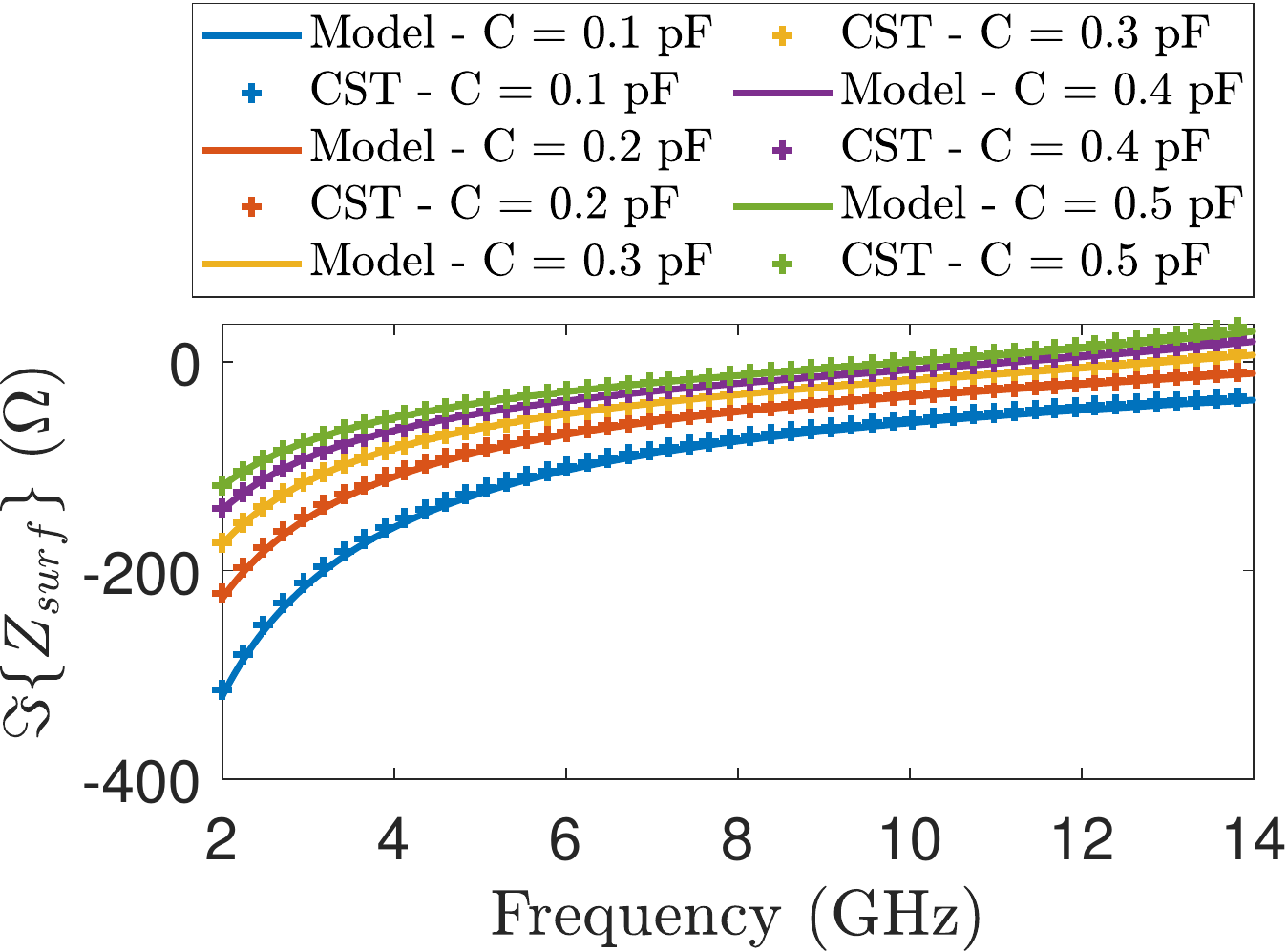}}
	\label{}\hfill        
	\caption{(a) Real and (b) imaginary part of the surface impedance $Z_{surf}$ as a function of frequency for different values of the varactor capacitance.}
	\label{fig_impedance_RIS} 
\end{figure}

\begin{figure}
	\centering
	\subfloat[\scalebox{0.9}{\textit{TE amplitude}}]{%
		\includegraphics[width=4cm,keepaspectratio]{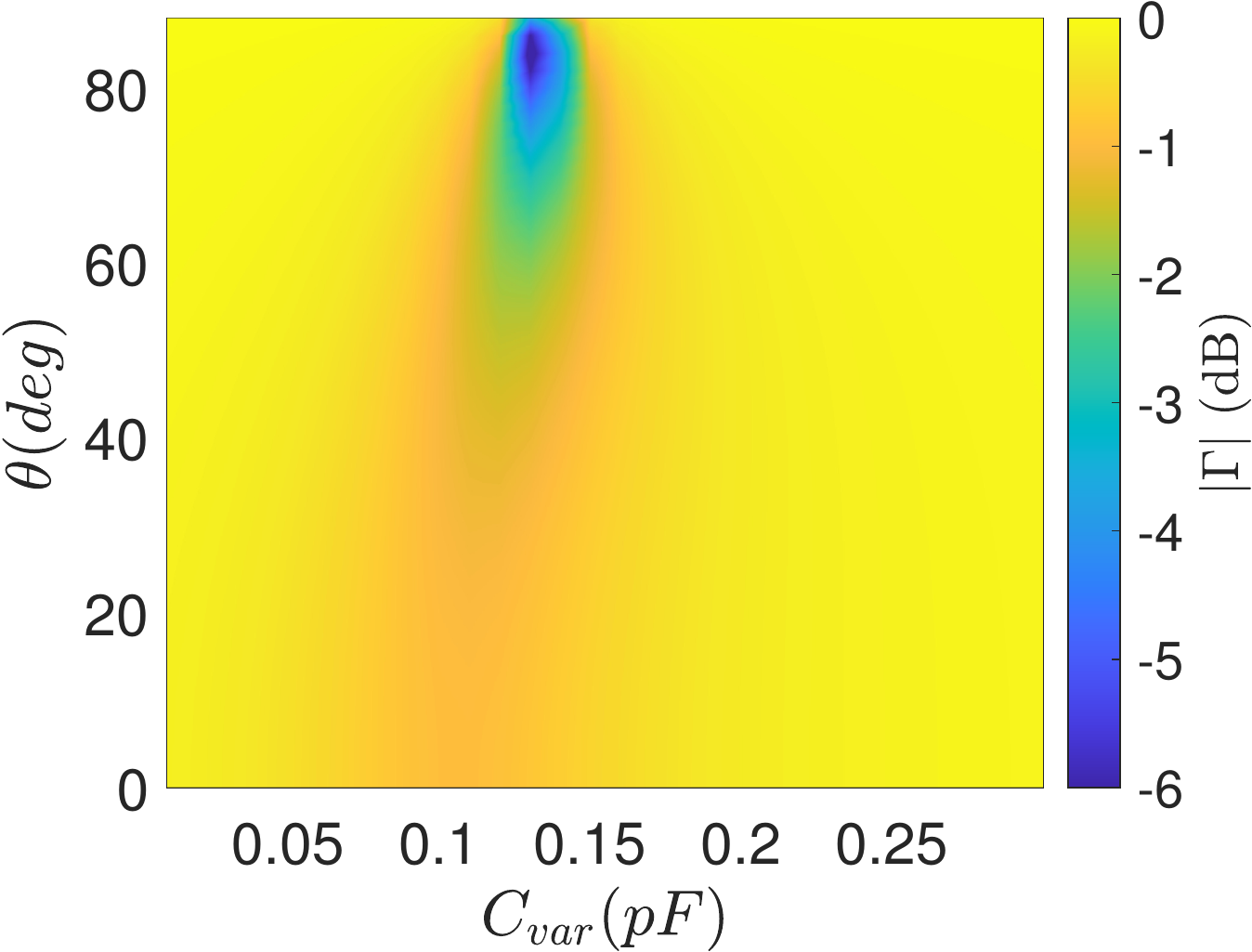}}
	\label{}\hfill        
	\subfloat[\scalebox{0.9}{\textit{TM amplitude}}]{%
		\includegraphics[width=4cm,keepaspectratio]{{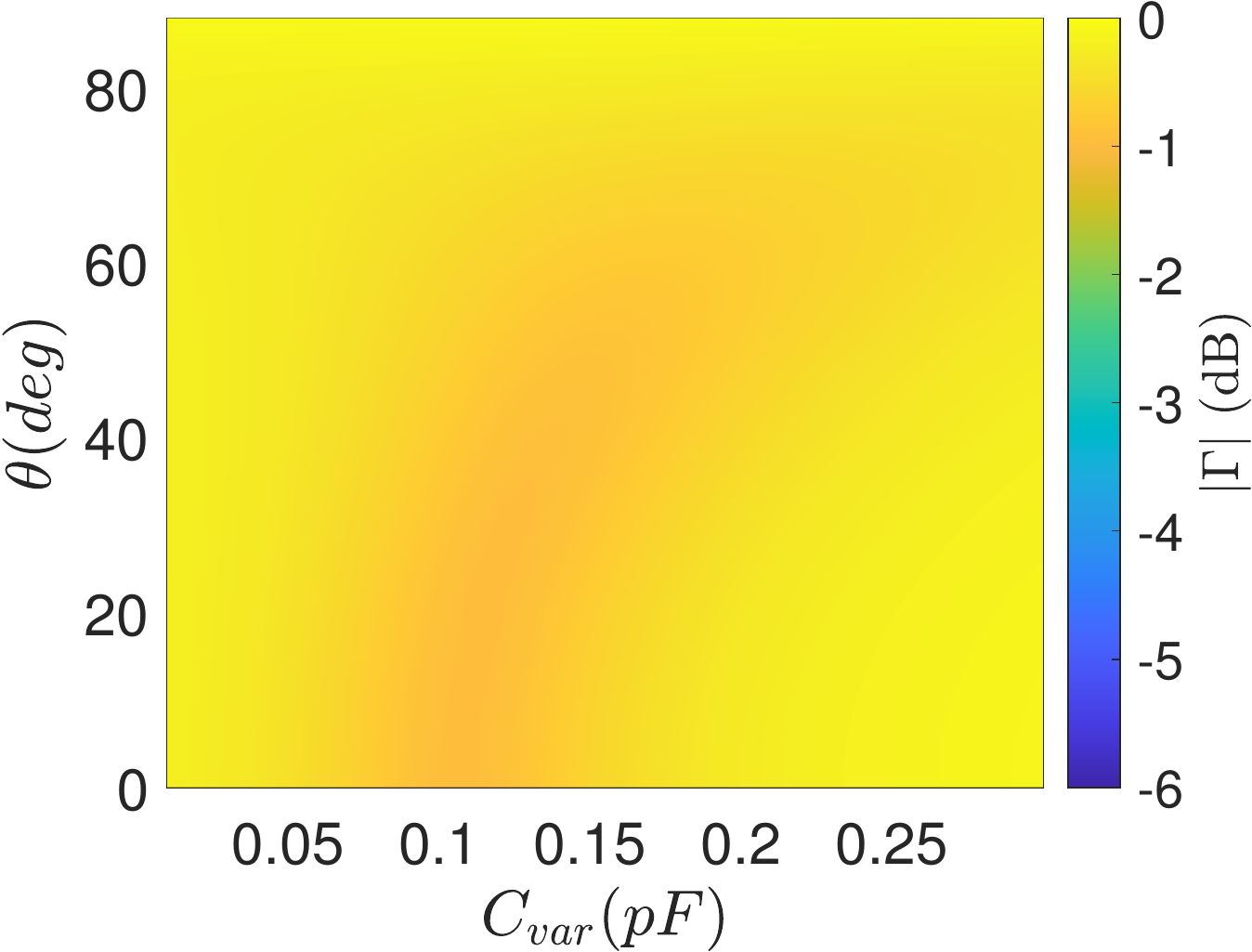}}}
	\label{}\hfill        
	\subfloat[\scalebox{0.9}{\textit{TE phase}}]{%
		\includegraphics[width=4cm,keepaspectratio]{{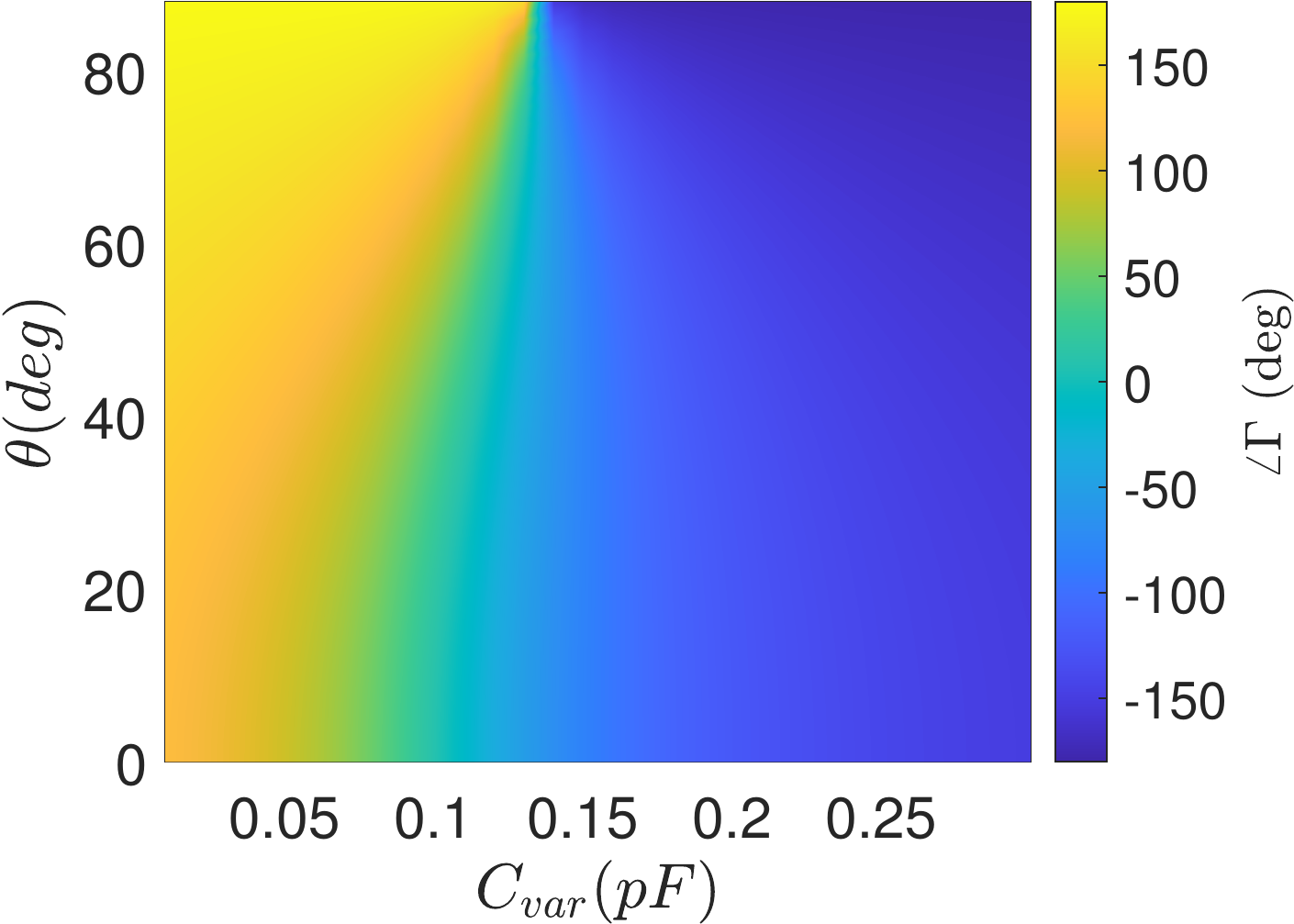}}}
	\label{}\hfill        
	\subfloat[\scalebox{0.9}{\textit{TM phase}}]{%
		\includegraphics[width=4cm,keepaspectratio]{{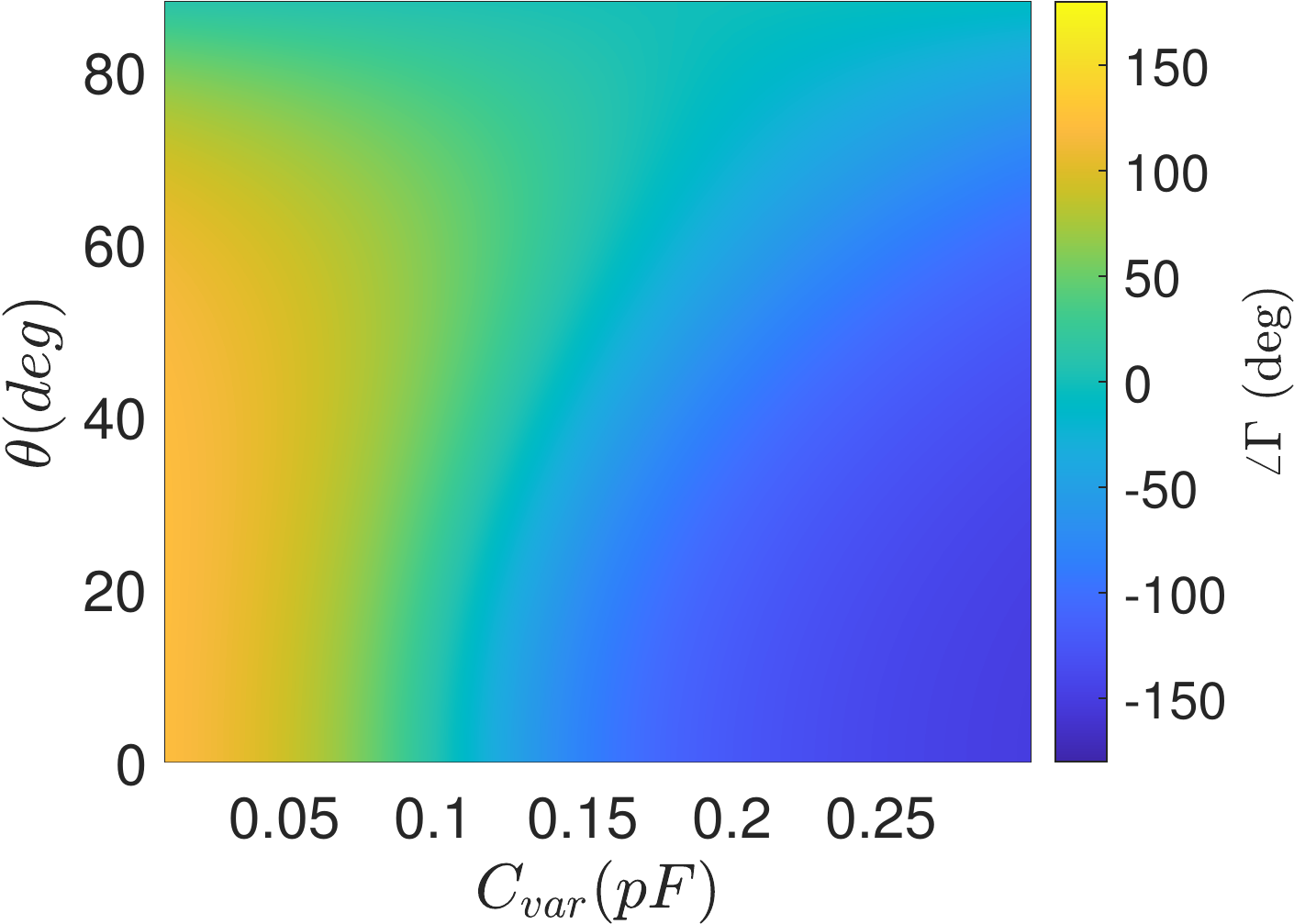}}}
	\label{}\hfill        
	\caption{(a) TE amplitude, (b) TM amplitude, (c) TE phase (d) TM phase. The geometrical and electrical parameters used for the RIS are the following: $D_x = D_y = D = 5$ mm, $w_x = 0.5$ mm,
		$\varepsilon_r = 4.4-j0.088$, $d = 1.2$ mm, $\sigma_{c} = 58.7 \times 10^6 S/m$ , $R_{var} = 0.5$   $\Omega$.}
	\label{fig_database} 
\end{figure}

The reflection phase of the surface controlled by varying the capacitance of the varactor diode is shown in Fig.~\ref{fig_reflection_phase_RIS}. The the proposed transmission line (TL) model is extremely accurate for both amplitude and phase reflection of the RIS. The amplitude and the phase response of the RIS are correlated. The minimum of the reflection amplitude appears close to the frequency where the phase is zero. If the structure is designed as a critically coupled resonator, the whole incoming EM energy is absorbed by the RIS and the structure acts as an electromagnetic absorber \cite{costa2012circuit}. In Fig.~\ref{fig_impedance_RIS}, the surface impedance of the periodic surface loaded with varactor diode is also shown. The behaviour of the impedance is shown as a function of the frequency for several values of the capacitance of the varactor. The impedance obtained by using the proposed model is compared with the impedance extracted from CST according to \cite{costa2009equivalent}. It is evident that the proposed model predicts very well the behaviour of the surface impedance till well above the resonance frequency of the RIS. It is also interesting to point out that the equivalent surface impedance turns from capacitive to inductive behaviour for relevant varactor capacitance values (i.e. $C_{var}=0.5$ pF). This is due to the fact that the varactor inductance plays an important role in the impedance behaviour. It has to be pointed out that both series inductance is considered in the CST model. The inductance of the lumped component is a distributed inductance created by the presence of the lumped capacitor between two neighbouring patches. Once that the accuracy of the model has been verified, the proposed simple model can be adopted to synthesize the phase properties of the RIS. To this aim, it is necessary to select some geometrical parameters for the RIS and then create a database of reflection coefficients as a function of diode varactor and incidence angle. Therefore, four matrix are needed: two for the reflection amplitude for TE and TM polarization and two for the reflection phase for TE and TM polarization. By using the proposed model, it is trivial to create a lookup table which links the capacitance state with the precise phase and amplitude value of the reflection coefficient for the required incidence angle and polarization. 
An example of bidimensional maps are reported in Fig.~\ref{fig_database} where the reflection coefficient as a function of the varactor capacitance and incident angle is shown through a color map. It is immediately clear how the phase value of the surface is characterized by a consistent phase variation for a certain capacitance value of the varactor. Also the reflection amplitude of the cell changes as a function of the angle both for TE and TM polarization. In the reported color maps, the ohmic losses in the periodic surface, the losses in the dielectric substrate ($\varepsilon_r=4.4-j0.088$) and the losses introduced by the series resistance of the varactor are considered ($R_{var}=0.5$ $\Omega$). For the case of patches array with $D_x = D_y = D$ and $w_x = w_y = w$, ohmic losses in the periodic surface are considered as \cite{costa2010analysis}:

\begin{equation} \label{eq_Gain_antenna} 
	R_{patch} = \left(\frac{D}{D-w}\right)^2R_s
\end{equation}

where $R_s=\frac{1}{\sigma_{c} \delta}$, $\omega$ is the angular frequency, $\delta$ is the skin depth of metal ($\delta=\frac{2}{\sqrt{\sigma_{c} \omega \mu_0}}$), $\mu_0$ is the magnetic permeability in a classical vacuum and the dielectric losses in the substrate are considered  while the resistance of the varactor is set to $1 \omega$. The conductivity of copper for the metallic surface is $\sigma_{c}=58.7 \times 10^6 S/m$.

\subsection{Reproducible Research}
The simulation results for can be reproduced using the code available at: 

	https://github.com/MicheleBorgese/Intelligent-Surfaces

\section{Link budget of RIS assisted communication} 
\label{sec_link} 

\begin{figure*}[hbt!]
	\centering
	\def\svgscale{0.75}
	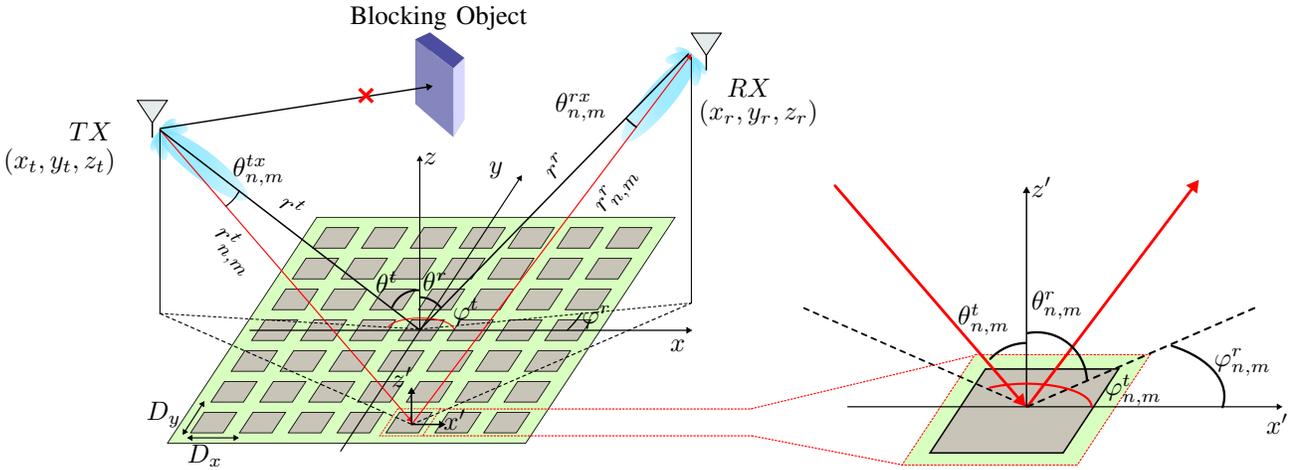
	\caption{Layout of the RIS-assisted non-line of sight communication scenario.}
	\label{fig_3D_link_RIS}
\end{figure*}

The link between the transmitter and receiver can be modelled as a classical backscattering communication system. The geometry of the communication path from the transmitted to the receiver passing though the RIS is shown in Fig.~\ref{fig_3D_link_RIS}. The RIS is placed in the $xy$-plane of a Cartesian coordinate system, and the geometric center of the RIS is aligned with the origin of the coordinate system. Let $M$ and $N$ denote the number of rows and columns of the regularly arranged unit cells of the RIS, respectively. The size of each unit cell is $D_x$ and $D_y$ along $x$ and $y$ axis respectively.
The received signal power in RIS-assisted wireless communications can be computed according to eq.(\ref{eq_link_budget}) which is shown in the next page. In this equation, $\lambda$ is the wavelength of the impinging plane wave, $ \Gamma_{m,n} $ is the reflection coefficient of the $(m,n)^{th}$ cell, which is evaluated by the proposed analytical approach described in section \ref{sec_RIS_model}. The pedexes $m,n$ denote the $m^{th}$ row and $n^{th}$ column of the programmable surface. The symbols $r^t_{m,n}$, and $r^r_{m,n}$ identify the distance of the $(m,n)^{th}$ unit cell from the transmitter (TX) and the receiver (RX), respectively. The elevation angle and azimuth angles of transmitter and receiver with respect to $(m,n)^{th}$ unit cell are identified by $ \theta^t_{m,n}$, $\varphi^t_{m,n}$, $ \theta^r_{m,n}$, $\varphi^r_{m,n}$. The polarization of the reflection coefficient can parallel or perpendicular depending on the polarization of the transmitting antenna.

\begin{figure*}[btp]
	\hrulefill
	\begin{equation}
		\begin{aligned}
			P_r = \frac{P_t}{(4\pi)^3} \lambda^2 \sum\limits_{m = 1}^M {\sum\limits_{n = 1}^N \frac{G_t(\theta^{tx}_{m,n}) G_r(\theta^{rx}_{m,n}) \sigma(\theta^t_{m,n},\theta^r_{m,n}) |\Gamma_{m,n}|^2 e^{-2j\left( {\angle{\Gamma_{m,n}}} \right)} e^{-j 2 k_0 (r^r_{m,n}+r^t_{m,n})}}{|r^t_{m,n}|^2 |r^r_{m,n}|^2}} 
		\end{aligned} \label{eq_link_budget}
	\end{equation}
\end{figure*}
    
As shown in Fig.~\ref{fig_3D_link_RIS}, the transmitter emits a signal towards the RIS with power $P_t$ through an antenna with power gain $G_t(\theta^{tx}_{n,m})$. The signal is reflected by the RIS and then received by the receiver with a gain $G_r(\theta^{rx}_{n,m})$. It is assumed that the polarization of the transmitter and receiver are always properly matched, even after the transmitted signal is reflected by the RIS. The radiation pattern of the transmitting and receiving antenna can be approximated through a $\cos^q$ function and thus the gain of the transmitting/receiving antenna to the specific unit cell of the RIS can be computed according to:

\begin{equation} \label{eq_Gain_antenna} 
	G(\theta) = \frac{4 \pi \cos^q(\theta)}{\displaystyle\int_0^{2 \pi} \displaystyle\int_0^{\pi/2} \cos^q(\theta)\, d\theta d\phi}
\end{equation}

The gain is only a function of the $\theta$ angle. The received power calculation in eq.(\ref{eq_link_budget}) is similar to the one followed in \cite{tang2020wireless}. Differently from \cite{tang2020wireless}, we used the concept of Radar Cross Section (RCS) \cite{balanis2012advanced}, which is more common for scattering problems (see also Appendix - \ref{Appendix-c}). Anyway, the results are consistent with \cite{tang2020wireless} since the maximum RCS of the unit cell can be approximated  as the gain of the unit cell multiplied by its area. 
The bistatic RCS of a unit cell can be computed according to bistatic RCS of a metallic plate. Assuming a parallel polarized incident plane wave (Fig.~\ref{fig_polarization}(a)) on the \textit{yz}-plane (this is commonly called $TM_x$ in scattering problems), the RCS in eq.(\ref{eq_link_budget}) can be written as follows:
  
\begin{multline} \label{eq_RCS_PEC_2D} 
	\sigma(\theta^t_{m,n},\theta^r_{m,n}) = 4 \pi \left(\frac{D_x D_y}{\lambda}\right)^2 \cos^2(\theta^t_{m,n}) \times \\ \times \left(\frac{\sin\left(\frac{kD_y}{2}\left(\sin\theta^r_{m,n}+\sin\theta^t_{m,n}\right)\right)}{\frac{kD_y}{2}\left(\sin\theta^r_{m,n}+\sin\theta^t_{m,n}\right)}\right)^2
\end{multline}

where $\theta^t_{m,n} \in [- \frac{\pi}{2},\frac{\pi}{2}]$ and $\theta^s_{m,n} \in [-\frac{\pi}{2},\frac{\pi}{2}]$. The RCS towards a generic direction is directly proportional to the physical area of the scattering object. If all unit cells provide the same phase response, the result approaches the one of a metallic plate of the same size of the RIS (see also Appendix - \ref{Appendix-d}).

In order to verify the importance of the TL model of the RIS also for oblique incidence, we present an example of a RIS-assisted communication between two antennas. Let us assume to have a transmitting antenna in the position $(-40,0,10)$ cm and a receiving antenna in the position $(20,0,20)$ cm. The RIS is composed by $30 \times 30$ cells with a size of $5$~mm. It is assumed that both the transmitting and receiving antennas are pointed towards the centrer of the surface and are linearly polarized with a field along \textit{y}-direction (this implies that the oblique incidence on the surface is TE polarized). The distance of the transmitter and receiver from the RIS are compact in order to maintain low the time needed to draw the two dimensional field maps. However, the proposed model can be applied to whatever distance from the RIS taking into account that the larger the RIS size, the larger the amount of energy deflected towards the receiver position. Indeed, the same principles used in the design of reflectarray or reflector antennas \cite{nayeri2013radiation, Borgi_reflectarray} hold for addressing which would be the most suitable panel size given a certain location of the transmitter. In the case of reflectarray antennas, the parameters known as illumination efficiency and spillover efficiency are used to determine the correct size of the reflector but in a communication scenario the location of the transmitter with respect to the RIS is usually unknown and the optimization of the RIS size is not needed. However, if the location of the transmitter with respect to the RIS is known, the RIS size may be optimized based on spillover efficiency and illumination efficiency. The ideal phases of the surface to maximize the power at the receiver location can be found analytically according to:

\begin{equation} \label{eq_optimal_phases} 
	\angle{\left(\Gamma_{m,n} \right)} = k_0 (r^r_{m,n}+r^t_{m,n})
\end{equation}

\begin{figure}
	\centering
	\subfloat[\scalebox{0.9}{\textit{ideal}}]{%
		\includegraphics[width=8.0cm,keepaspectratio]{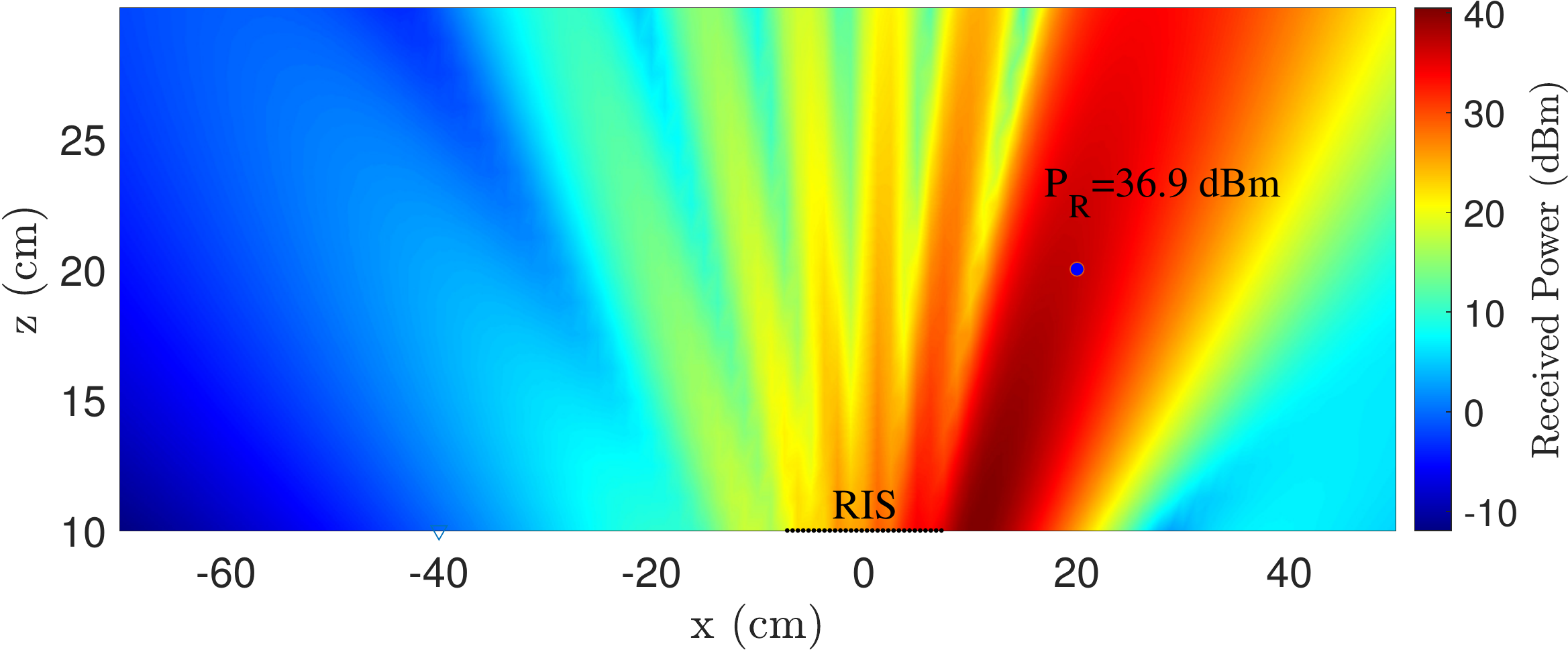}}
		\label{}\hfill        
	\subfloat[\scalebox{0.9}{\textit{normal incidence model}}]{%
		\includegraphics[width=8.0cm,keepaspectratio]{{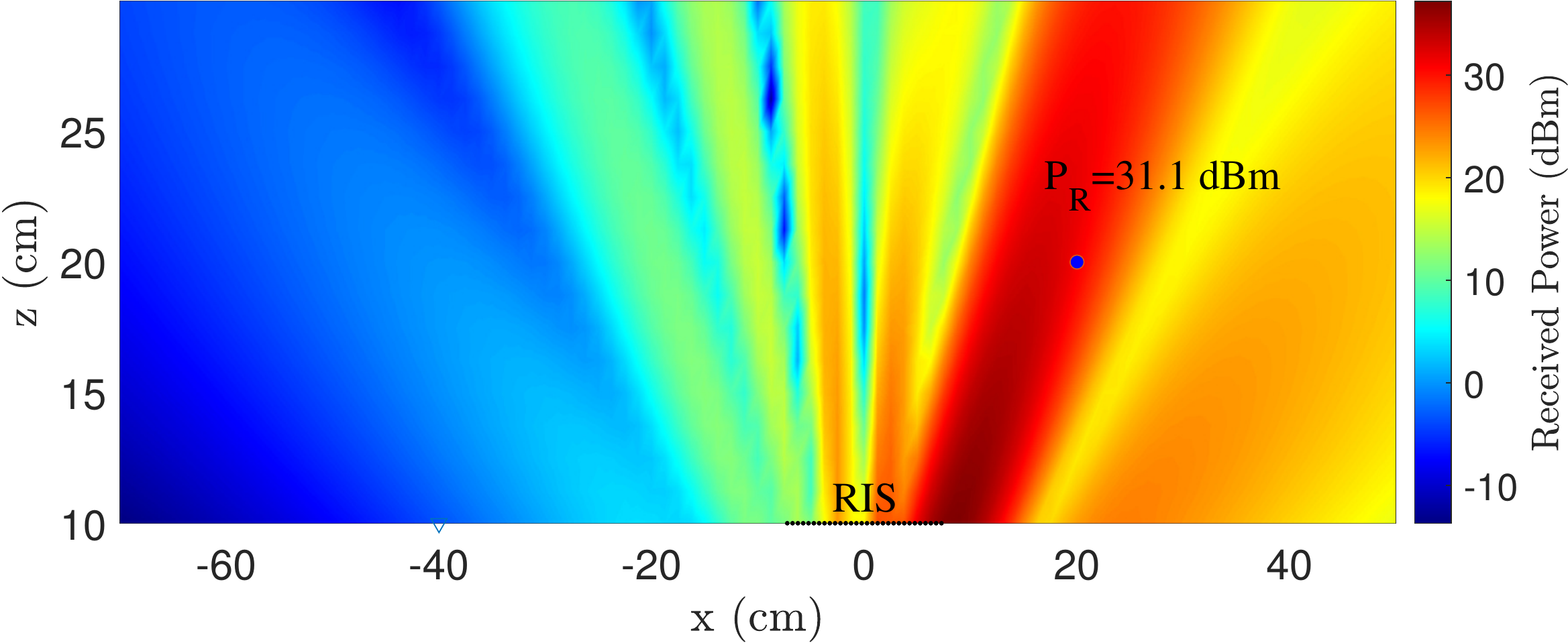}}}
	\label{}\hfill        
	\subfloat[\scalebox{0.9}{\textit{proposed model}}]{%
		\includegraphics[width=8.0cm,keepaspectratio]{{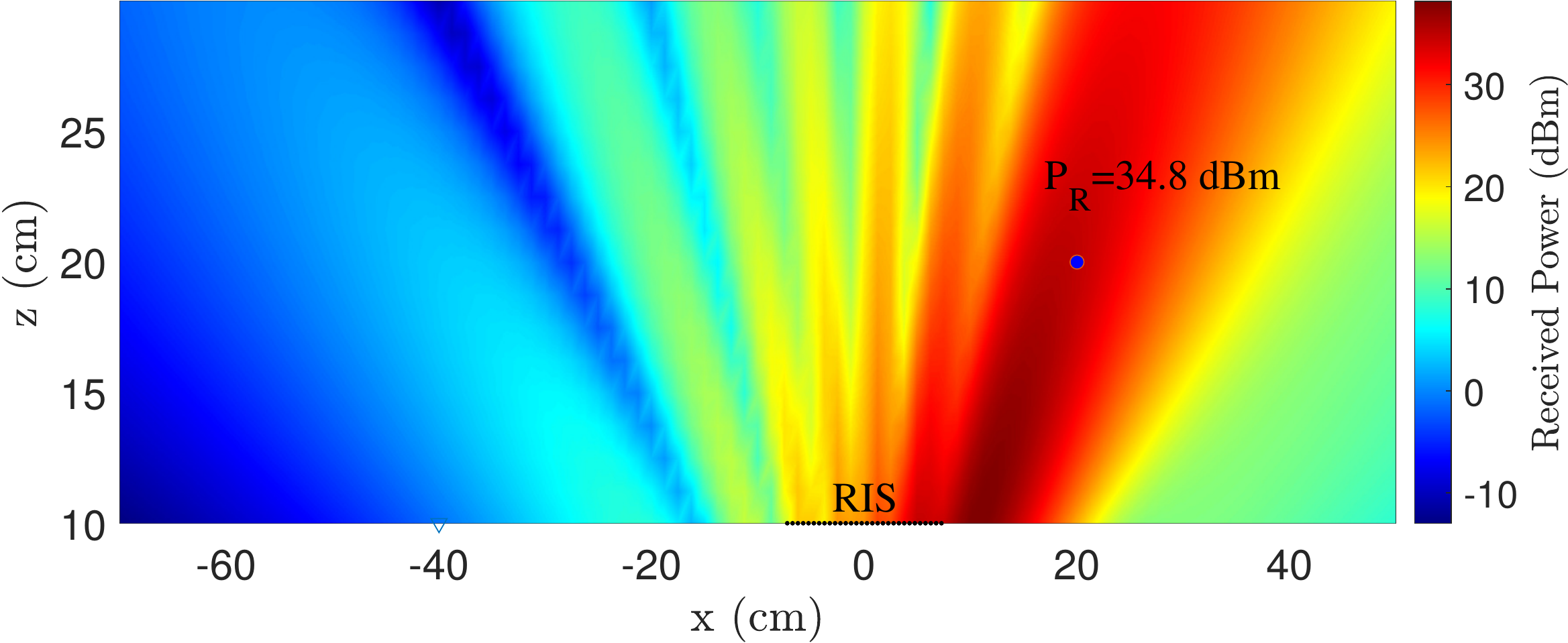}}}
	\label{}\hfill             
	\caption{Received power ($P_r$) on the \textit{xz}-plane at 8 GHz: (a) ideal, (b) normal incidence model, (c) proposed model. Both the normal incidence model and the proposed (oblique incidence) model include the effect of reflection losses due to non-idealities of the RIS (ohmic loss, substrate loss, varactor loss) but the oblique incidence model also consider the specific angle of incidence for each unit cell for deriving the optimal varactor state. }
	\label{fig_plot_PR_2D} 
\end{figure}

\begin{figure}
	\centering
	\subfloat[][\scalebox{0.9}{\textit{normal incidence model}}]{%
		\includegraphics[width=4cm,keepaspectratio]{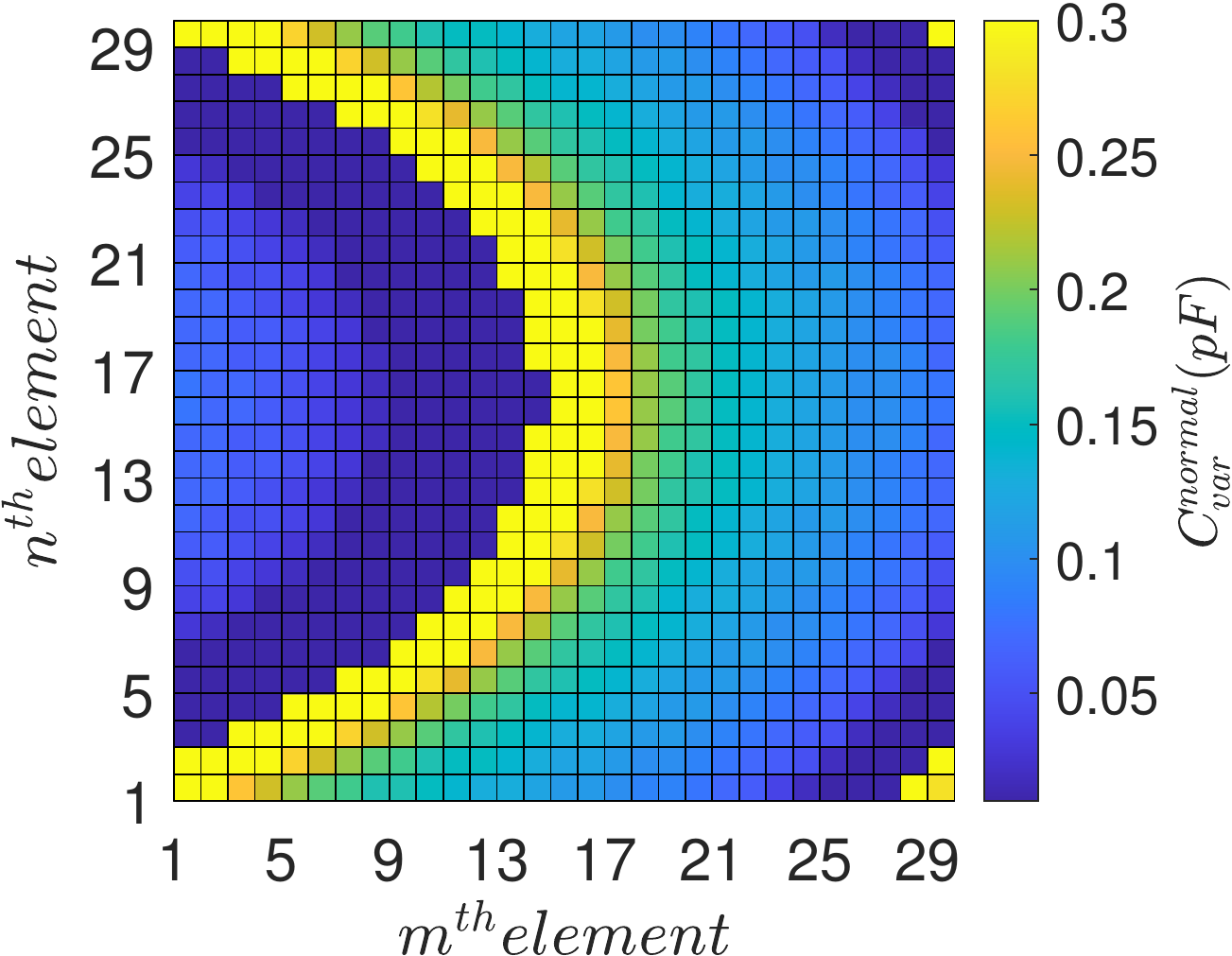}}
	\label{}\hfill       
	\subfloat[][\textit{\scalebox{0.9}{\textit{proposed model}}}]{%
		\includegraphics[width=4cm,keepaspectratio]{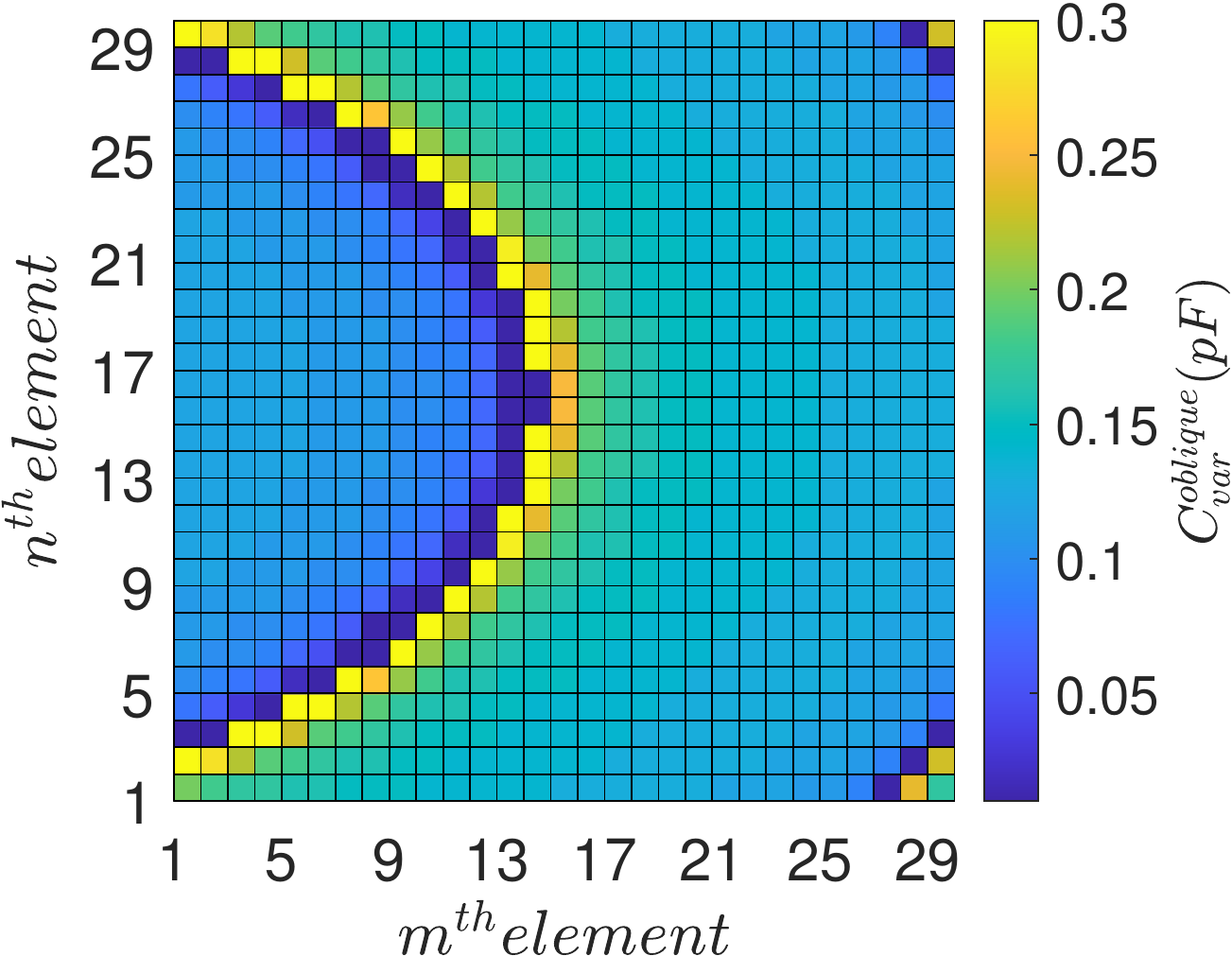}}
	\label{}\hfill       
	\subfloat[][\textit{\scalebox{0.9}{\textit{Percentage error}}}]{%
		\includegraphics[width=4cm,keepaspectratio]{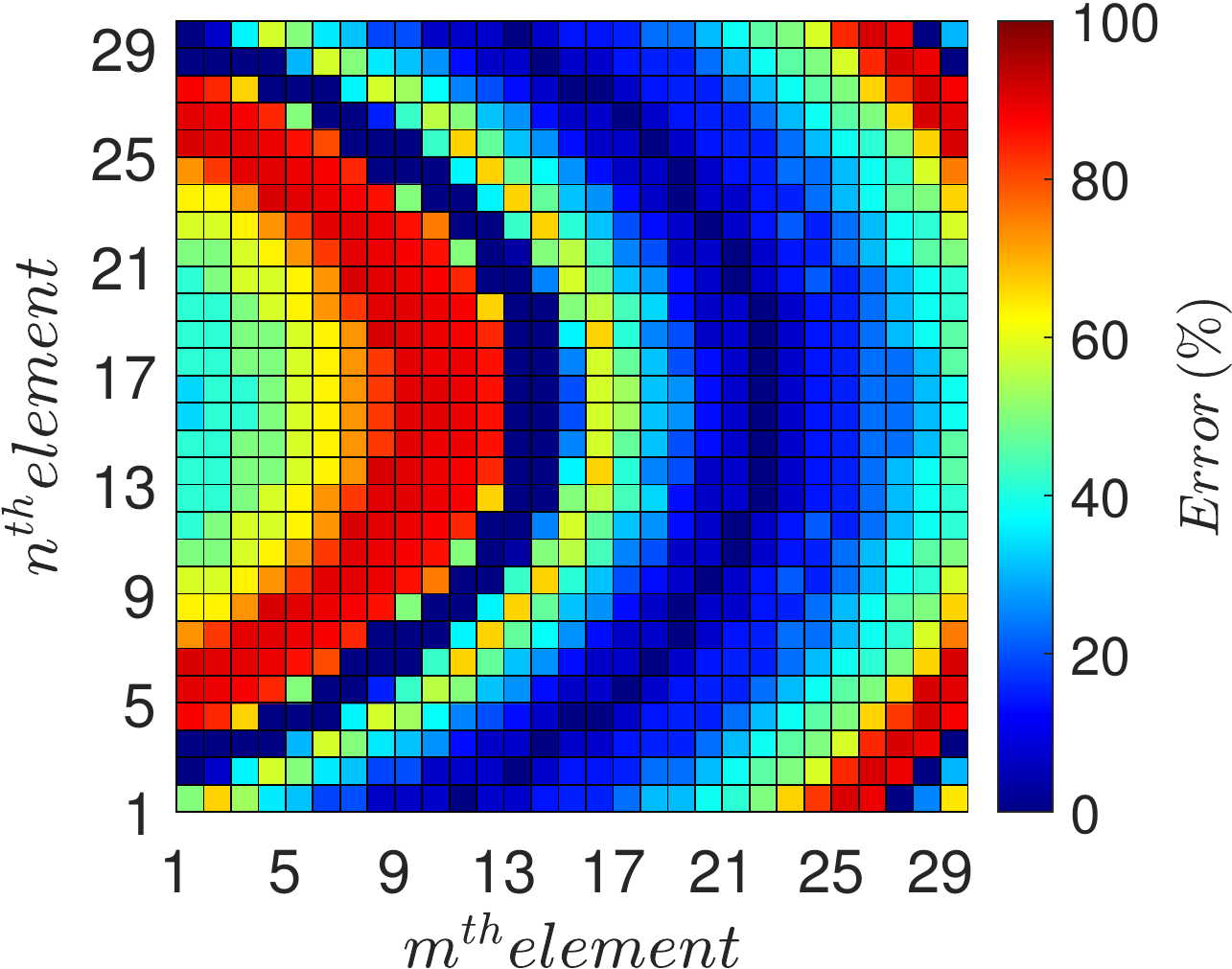}}
	\label{}\hfill       
	\caption{Capacitances optimized with (a) normal incidence ($C_{var}^{normal}$) model and with (b) oblique incidence ($C_{var}^{oblique}$) model, (c) percentage error of the capacitance computed with normal incidence model: $Error (\%)=|(C_{var}^{oblique}-C_{var}^{normal})/(C_{var}^{oblique})| \times 100$.}
	\label{fig_capacitances_of_the_cells} 
\end{figure}

However, in order to design an actual RIS which physically implements the behaviour of the ideal surface, the ideal phases found according to (\ref{eq_optimal_phases}) should be translated into the geometrical parameters of the RIS (shape of the unit cell, periodicity, dielectric thickness and dielectric permittivity) and the capacitance values provided by the varactors connected in each unit cell. This process can be advantageously accomplished through the model presented in section \ref{sec_Model}. Two cases are possible: the incidence angle with respect to each unit cell is neglected and all the phases are computed at normal incidence; alternatively, the incident angle for each unit cell is considered and the exact capacitance value needed for achieving the desired phase at the incidence angle specific for the unit cell is computed. It is evident that, while the optimal phases can be synthesized by properly selecting the geometrical parameters of the unit cell and the capacitance state, the ideal amplitudes cannot be synthesized since a realistic surface realization is accompanied by some intrinsic absorption losses of the surface.  A 2D plot of the power radiated in the plane parallel to both transmitting and receiving antenna is reported in Fig.~\ref{fig_plot_PR_2D} for three cases: ideal, normal incidence phases, oblique incidence phases. As is evident, the optimal situation is the one with the ideal phase values and the ideal amplitude values (perfect reflection). If the phases are optimized considering normal incidence for all elements, a considerable drop of the received power is obtained both because of the non idealities due to the absorption phenomena and more importantly for the phase detuning. The third approach, the proposed one, which relies on the oblique incidence model instead allows to limit the power drop with respect to the ideal situation since the phase detuning due to the spatial dispersion (oblique incidence effect) of the surface is eliminated. In this example a power gain of 3.9 dB is achieved by using the proposed approach compared to the normal incidence approach. In general, the achieved power gain considering the oblique incidence can be lower or even much higher if the impinging EM waves come from grazing angles with respect to the normal to the surface. The only case in which the normal incidence approach can guarantee the same performance of the proposed approach is the particular situation where the EM wave impinges at normal incidence. The optimized values of the varactor capacitances are reported in Fig.~\ref{fig_capacitances_of_the_cells}. Both the cases of the normal incidence model and the proposed oblique incidence model are shown. The same figure reports also the percentage error of the capacitances synthesized by neglecting the dependence on incidence angle.    
The proposed method can be easily adopted in all multi-user beamforming approaches as for replacing the ideal phase and amplitude responses of the RIS which is usually employed. The phase values are usually hypothesized in the range $[0, 2\pi)$ \cite{wu2019beamforming} (sometimes they are discretized in a finite number of states) and the amplitude are considered equal to 1. This hypothesis is actually not viable because of hardware limitations. For instance, as  reported in Fig.~\ref{fig_reflection_phase_RIS}(b), with the selected geometrical parameters, the range of phase change in the range ($-170^\circ$, $110^\circ$) while the amplitude changes depending of the varactor states and is always lower than 0 dB because of intrinsic losses. Moreover, the phase values change also as a function of the incidence angle of the impinging wave as shown in Fig.~\ref{fig_database}.

\section{Conclusion}
\label{sec_conclusion} 
An accurate analytical model for the synthesis of reconfigurable intelligent surfaces has been presented. The model is based on an analytical representation of the periodic surface via homogenized impedance and a transmission line circuit approach. The RIS reflection coefficient is controlled through varactors connecting periodic patches which are included in the model through an RLC series impedance connected in parallel to the surface impedance of the periodic surface. The varactors can be programmed through a biasing network to adjust locally the reflection coefficient phase of the RIS in order to achieve the desired beam forming and the corresponding maximization of the power on the receiver position. It has been shown that the use of the proposed model, which is capable of catching the reflection coefficient behaviour both at normal and oblique incidence for both for TE and TM polarizations, provide a consistent power gain in the design of RIS-assisted communications.

\appendices
\section{Floquet theorem for periodic structures}
\label{Appendix-a}

\begin{figure}
	\centering
	\def\svgscale{1.5}
	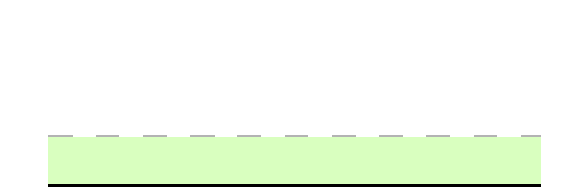
	\caption{Bidimensional representation of a periodic Artificial Impedance Surface.}
	\label{fig_AIS_floquet}
\end{figure}

The reconfigurable intelligent surface is basically a periodic structure perturbed by the local change of active components. The analysis of the surface can be efficiently performed by analysing a single unit cell under the local periodicity approach and thus by exploiting the Floquet theorem. This approach allows to consider the effect of neighbouring cells in the reflection coefficient and consequently the effect of the mutual coupling. For simplicity, we describe a monodimensional periodic surface lying on the \textit{xz}-plane and with a periodicity along \textit{x}-direction as shown in Fig.~\ref{fig_AIS_floquet}. A similar approach is valid for two-dimensional periodic surfaces.

In presence of a periodic structure as the one reported in Fig.~\ref{fig_AIS_floquet}, any electromagnetic field component can be described as:

\begin{equation} \label{floquet} 
	\phi(x+D_x,y,z) = e^{-jk_{x0} D_x} \phi(x,y,z)
\end{equation}

The exponential factor indicates the complex phase shift between neighbouring cells $(k_{x0}=\beta-j\alpha)$, implying that the field components differ only for a phase term. This relation is called Floquet Theorem \cite{itoh2004periodic}. The field component satisfying eq.(\ref{floquet}) can be written described as:

\begin{equation} \label{floquet2} 
	\phi(x,y,z) = e^{-jk_{x0} D_x} P(x,y,z)
\end{equation}

where $P$ is a periodic function $\left( P(x,y,z) = P(x+D_x,y,z) \right)$ which can be therefore expanded in a Fourier series:

\begin{equation} \label{floquet3} 
	P(x,y,z) = \sum\limits_{n = -\infty}^\infty a_n(z,y) e^{-j \frac{2 \pi n}{D_x} x}
\end{equation}
   
where $a_n$ is a functions that describes the field variation in \textit{yz}-plane. By using eq.(\ref{floquet2}) and eq.(\ref{floquet3}), the field can be finally written as:

\begin{equation} \label{floquet4} 
	\phi(x,y,z) = \sum\limits_{n = -\infty}^\infty a_n(x,y) e^{-j k_{xn} x}
\end{equation}

where 

\begin{equation} \label{floquet5} 
	k_{xn}=k_{x0}+\frac{2 \pi}{D_x} n,  \:\:\:\:  with \:\:\:\:  n=0,\pm1,\pm2...
\end{equation}

The terms of the summation in eq.(\ref{floquet4}) are usually referred as \textit{Floquet spatial harmonics}. The Floquet harmonics consist of a set of plane waves which can be in propagation (the propagation constant $\beta$ is purely real) or in cut-off (the propagation constant $\beta$ is purely imaginary) but all of them contribute in the determination of the local field distribution.

\section{Analysis of a freestanding impedance surface}
\label{Appendix-b}

\begin{figure}
	\centering
	\def\svgscale{0.95}
	\subfloat[]{%
		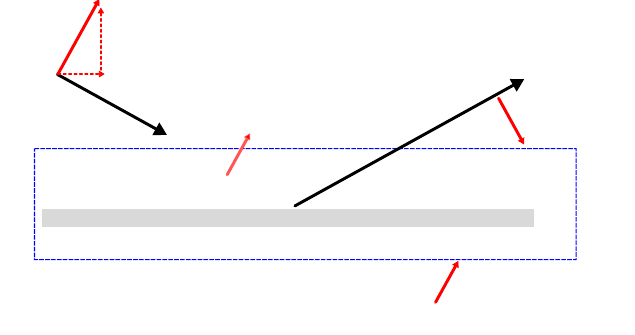}
\hfill
	\subfloat[][\scalebox{0.9}{\textit{}}]{%
		\includegraphics[width=2.5cm,keepaspectratio]{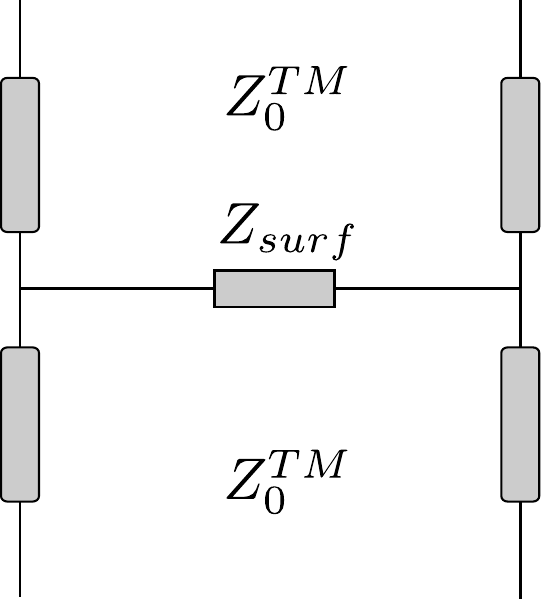}}
	\label{}\hfill
	\caption{(a) Uniform $TM_x$ plane wave incident on an impedance surface characterized by a surface impedance $Z_{surf}$; (b) Transmission line model of a freestanding metasurface.}
	\label{fig_impedance_surface_scattering}
\end{figure}

Let us consider a TM impinging electromagnetic wave towards a uniform sheet characterized by a complex surface impedance $Z_{surf}$ as shown in Fig.~\ref{fig_impedance_surface_scattering}(a). The impinging electric and magnetic fields can be written as follows:

\begin{equation} \label{Ei} 
	\vec E^i = E_0  \left(\hat{i}_y \cos \theta^i+\hat{i}_z \sin\theta^i \right)  e^{-j\beta(y \sin\theta^i - z \cos \theta^i)}
\end{equation}

\begin{equation} \label{Hi} 
	\vec H^i = \frac{E_0}{\zeta_0}  \hat{i}_x e^{-j\beta(y \sin\theta^i - z \cos \theta^i)}
\end{equation}

Indicating with $\Gamma^{TM}$ the reflection coefficient of the surface for TM incidence, the scattered fields have the following form:

\begin{equation} \label{Es} 
	\vec E^s = \Gamma^{TM} E_0  \left(\hat{i}_y \cos \theta^i+\hat{i}_z \sin\theta^i \right)  e^{-j\beta(y \sin\theta^i + z \cos \theta^i)}
\end{equation}

\begin{equation} \label{Hs} 
	\vec H^s = \frac{-\Gamma^{TM} E_0}{\zeta_0} \; \hat{i}_x e^{-j\beta(y \sin\theta^i + z \cos \theta^i)}
\end{equation}

The uniform sheet, characterized by an isotropic electric surface impedance $Z_{surf}$, supports the propagation of a average surface current $J_{av}$, therefore:

\begin{equation} \label{eq_Et} 
	\hat{i}_n \times \vec E^{+} =  J_{av} \; Z_{surf}
\end{equation}

The surface current is given by the discontinuity of the magnetic field at the interface. Therefore the boundary condition in (\ref{eq_Et}) can be also written as:

\begin{equation} \label{eq_GSTC} 
	\hat{i}_n \times \vec E^{+} = \hat{i}_n \times \left( Z_{surf} \cdot \left(\hat{i}_n \times \left( \vec{H}^+ - \vec{H}^-  \right) \right)\right)
\end{equation}
where $J_{av}= \hat{i}_n \times \vec{H}^+$ and $\hat{i}_n $ is the (outward) unit vector normal to the metasurface.  The subscript $+$ and $-$ indicate that the field is evaluated close to the surface from the incident side and transmission side, respectively \cite{francavilla2015numerical}. If the surface is anisotropic the surface impedance becomes a tensor \cite{francavilla2015numerical,borgese2020simple}. The tangential electric field is instead continuous at the interface: 

\begin{equation} \label{eq_Et_continuity} 
	\hat{i}_n \times \vec E^{+} =  \hat{i}_n \times \vec E^{-} 
\end{equation}
\\

Representing with $\tau^{TM}$ the transmission coefficient of the surface for TM incidence and applying the boundary conditions at $z=0$ for the tangential components of the field (continuity of the tangential components of the fields and impedance boundary condition given in (\ref{eq_GSTC}), the following relations are valid:

\begin{equation} \label{E_continuity} 
1+\Gamma^{TM} =\tau^{TM}
\end{equation}

\begin{equation} \label{H_discontinuity} 
-\tau^{TM} \cos \theta^i =Z_{surf} \left[ \tau^{TM} \frac{1}{\zeta_0}-  \left( \frac{1}{\zeta_0} - \Gamma^{TM} \frac{1}{\zeta_0} \right) \right]
\end{equation}

Substituting (\ref{E_continuity}) into (\ref{H_discontinuity}), and solving for $\Gamma^{TM}$ and $\tau^{TM}$, we obtaine the reflection and transmission coefficient  of the freestanding metasurface:

\begin{equation} \label{tau_TM} 
\tau^{TM}=\frac{2Z_{surf}}{2Z_{surf}+\zeta_0 \cos \theta^i}
\end{equation}

\begin{equation} \label{tau_TM} 
\Gamma^{TM}=\frac{-\zeta_0 \cos \theta^i}{2Z_{surf}+\zeta_0 \cos \theta^i}
\end{equation}

The same result can be obtained by using the transmission line model depicted in  Fig.~\ref{fig_impedance_surface_scattering}(b). Indeed, the reflection coefficient can be derived by:

\begin{equation} \label{eq_Zv} 
	\Gamma^{TM}=\frac{Z_{v} - \zeta_0^{TM}}{Z_{v}+\zeta_0^{TM}}
\end{equation}

where $\zeta_0^{TM}=\zeta_0 \cos \theta^i$ and $Z_v$:

\begin{equation} \label{eq_Zv} 
	Z_{v}=\frac{Z_{surf} \; \zeta_0^{TM}}{Z_{surf}+\zeta_0^{TM}}
\end{equation}

An effective approach to model the behavior of periodic surfaces is to establish an analogy with lumped filters. The surface impedance can be also represented as a single impedance having an LC response \cite{costa2014overview, costametamaterials}.

\section{RCS of a Conducting Flat Plate}
\label{Appendix-c}
 
\begin{figure}
	\centering
	\subfloat[][\scalebox{0.9}{\textit{}}]{%
		\includegraphics[width=4.5cm,keepaspectratio]{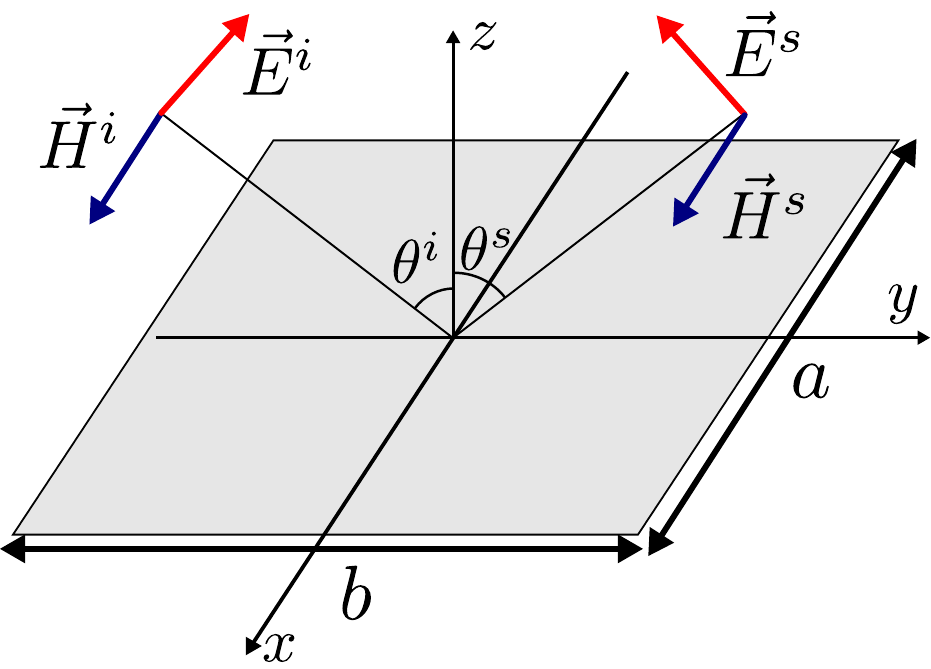}}
	\label{}\hfill  
	\subfloat[][\scalebox{0.9}{\textit{}}]{%
		\includegraphics[width=4.1cm,keepaspectratio]{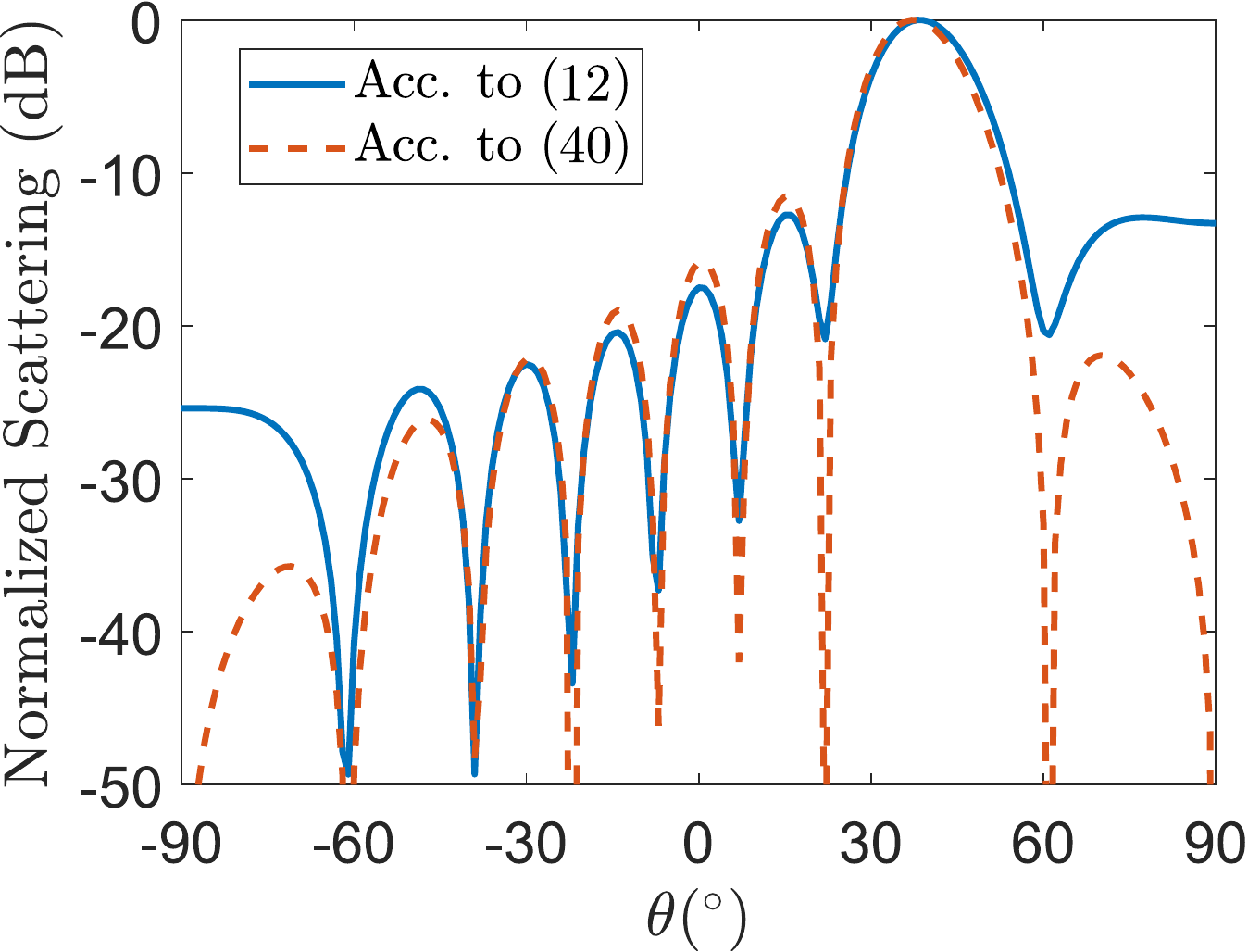}}
	\label{}\hfill  
	\caption{(a) Uniform $TM_x$ plane wave incident on a rectangular conducting plate; (b) far-field scattering of square a metallic plate ($a = b = 15$ cm) of with an incident angle of $\theta^i = \theta^s = 38.6 ^\circ$ computed with the double summation approach (according to (\ref{eq_link_budget})) considering the metallic surface composed by $ 30 \times 30$ perfectly conductive unit cells and with the RCS relation for a flat plate (according to (\ref{eq_RCS_PEC_2D})). The scattered field of each curve is normalized to the maximum value which is reached when $\theta^i = \theta^s$ (bistatic scattering).}
	\label{fig_Plate_scattering} 
\end{figure}

The RCS of a plane wave on a flat plate can be computed both for TE and TM polarized waves. The derivation are pretty similar for the two polarizations but what is changing are the tangential components of the surface current excited on the metal plate. We present the TM case but similar considerations hold for TE waves. 

Let us assume to have a parallel polarized ($TM_x$) plane wave on the \textit{yz}-plane impinging with an angle $\theta_i$ upon a rectangular electric perfectly conducting flat plate of dimensions $a$ and $b$ as shown in Fig.~\ref{fig_Plate_scattering}(a). The incident electric and magnetic fields can be written as follow:

\begin{equation} \label{eq_X_balanis} 
	\vec E^i = E_0  \left(\hat{i}_y \cos \theta^i+\hat{i}_z \sin\theta^i \right)  e^{-j\beta(y \sin\theta^i - z \cos \theta^i)}
\end{equation}

\begin{equation} \label{eq_X_balanis} 
    \vec H^i = \frac{E_0}{\zeta_0}  \hat{i}_x e^{-j\beta(y \sin\theta^i - z \cos \theta^i)}
\end{equation}

The surface current induced in the metallic place reads:

\begin{multline} \label{eq_balanis_Js} 
	\vec J_s = \hat{i}_z \times \left(\vec H^i+\vec H^s \right) |_{z=0} =2 \hat{i}_y \vec H^i|_{z=0} = \\ =\hat{i}_y \; 2 \frac{E_0}{\zeta_0}e^{-j\beta y \sin\theta^i}
\end{multline}

The surface current for the reported case is only in the \textit{y}-direction and it is different from zero only in the region occupied by the metallic surface. The scattered field can be computed by evaluating the integral of the flat current on the finite sized surface. The finite integral has the form of a $sinc$ function. Once computed the $E_\theta$ and the $E_\phi$ component of the scattered field  \cite{balanis2012advanced}, the radar cross section of the metallic plate can be computed in three dimensional space can be written as follow:

\begin{multline} \label{eq_RCS_PEC}  
	\sigma(\theta^i,\theta^s,\varphi^s) = 4 \pi \left(\frac{a b}{\lambda}\right)^2 \\
	 \left(\cos^2\theta^s \sin^2\varphi^s + \cos^2\varphi^s\right)  
	\left(\frac{\sin X}{X}\right)^2 \left(\frac{\sin Y}{Y}\right)^2 
\end{multline}

where $X$ and $Y$ are defined as:

\begin{equation} \label{eq_X_balanis} 
	X = \frac{ka}{2}\sin\theta^s \cos\varphi^s
\end{equation}

\begin{equation} \label{eq_Y_balanis} 
	Y = \frac{kb}{2}\left(\sin\theta^s \sin\varphi^s -\sin\varphi^i\right)
\end{equation}

If we restrict the analysis to the plane of incidence, yz plane, the expression of the RCS simplifies as \cite{ozdogan2019intelligent}:

\begin{multline} \label{eq_RCS_PEC_2D} 
	\sigma(\theta^i,\theta^s) = 4 \pi \left(\frac{a b}{\lambda}\right)^2 \cos^2(\theta^i) \\ \left( \frac{\sin\left(\frac{kb}{2}\left(\sin\theta^s-\sin\theta^i\right)\right)}{\frac{kb}{2}\left(\sin\theta^s-\sin\theta^i\right)} \right)^2
\end{multline}

The far-field scattering of square a metallic plate computed with the double summation approach (according to (\ref{eq_link_budget})) considering the metallic surface composed by $ 30 \times 30$ perfectly conductive unit cells and with the RCS relation for a flat plate (according to (\ref{eq_RCS_PEC_2D})) is shown in Fig.~\ref{fig_Plate_scattering}(b).

\section{Power received with a metallic plate in far field}
\label{Appendix-d}
  
The expression for the calculation of the power reflected by the RIS can be simplified to an analytical form in case that the transmitter and the received are far from the surface and the reflection coefficient of all unit cells are identical. Indeed, in this case, the surface just introduce a phase delay to the whole impinging rays. The Perfect Electric Conductor (PEC) case is a particular case of uniform surface where all the elements are metallic and thus they are characterized by a reflection coefficient equal to -1. In this case the power scattered pattern should agree with the canonical obtained by using the classical expression for radar cross section of a metallic plate \cite{balanis2012advanced}. Indeed, in case of $N \times M$ identical metallic elements, the double summation in eq.~(\ref{eq_link_budget}) leads to:
             				
\begin{equation} \label{eq_received_power_PEC} 
	P_r = \frac{P_t}{(4\pi)^3} \lambda^2 (MN)^2 \frac{G_t(\theta^{tx}) G_r(\theta^{rx}) \sigma(\theta^t,\theta^r)}{|r^t|^2 |r^r|^2}
\end{equation}

By substituting the closed form expression for the unit cells RCS, we obtain:

\begin{multline} \label{eq_received_power_PEC2} 
	P_r = \frac{P_t}{(4\pi)^2} (MN D_x D_y)^2 \frac{G_t(\theta^{tx}) G_r(\theta^{rx})}{|r^t|^2 |r^r|^2} \cos^2(\theta^t) \\ \left(\frac{\sin\left( \frac{kD_y}{2}\left(\sin\theta^r-\sin\theta^t\right)\right)}{\frac{kD_y}{2}\left(\sin\theta^r-\sin\theta^t\right)}\right)^2
\end{multline}

The power is maximized on if the incident and scattered angle are the same but with a 180° rotation of the azimuth angle (bistatic case). Moreover, the relation highlight that, once fixed the gain of the transmitter and receiver, the power available at the receiver only depends on the size of the RIS and on the transmitter-surface distance distance, $r^t$ and surface-receiver distance, $r^r$ and not on the frequency. In the bistatic case, the received power can be simplified as:

\begin{equation} \label{eq_received_power_PEC2} 
	P_r = \frac{P_t}{(4\pi)^2} (D_x^{RIS} D_y^{RIS})^2 \frac{G_t(\theta^{tx}) G_r(\theta^{rx})}{|r^t|^2 |r^r|^2} \cos^2(\theta^t)
\end{equation}
\\
where $D_x^{RIS}=MD_x$ and $D_y^{RIS}=ND_y$.

\bibliographystyle{IEEEtran}
\bibliography{references}

\end{document}

%% file: Figure/TE_TM_inc/TE_TM_inc_parallel.pdf_tex
\begingroup%
  \makeatletter%
  \providecommand\color[2][]{%
    \errmessage{(Inkscape) Color is used for the text in Inkscape, but the package 'color.sty' is not loaded}%
    \renewcommand\color[2][]{}%
  }%
  \providecommand\transparent[1]{%
    \errmessage{(Inkscape) Transparency is used (non-zero) for the text in Inkscape, but the package 'transparent.sty' is not loaded}%
    \renewcommand\transparent[1]{}%
  }%
  \providecommand\rotatebox[2]{#2}%
  \newcommand*\fsize{\dimexpr\f@size pt\relax}%
  \newcommand*\lineheight[1]{\fontsize{\fsize}{#1\fsize}\selectfont}%
  \ifx\svgwidth\undefined%
    \setlength{\unitlength}{165.2703509bp}%
    \ifx\svgscale\undefined%
      \relax%
    \else%
      \setlength{\unitlength}{\unitlength * \real{\svgscale}}%
    \fi%
  \else%
    \setlength{\unitlength}{\svgwidth}%
  \fi%
  \global\let\svgwidth\undefined%
  \global\let\svgscale\undefined%
  \makeatother%
  \begin{picture}(1,0.45921105)%
    \lineheight{1}%
    \setlength\tabcolsep{0pt}%
    \put(0,0){\includegraphics[width=\unitlength,page=1]{TE_TM_inc_parallel.pdf}}%
    \put(0.22454515,0.3518438){\color[rgb]{0,0,0}\makebox(0,0)[lt]{\lineheight{1.25}\smash{\begin{tabular}[t]{l}$\vec{E^i_z}$\end{tabular}}}}%
    \put(0.21903957,0.26241375){\color[rgb]{0,0,0}\makebox(0,0)[lt]{\lineheight{1.25}\smash{\begin{tabular}[t]{l}$\vec{E^i_x}$\end{tabular}}}}%
    \put(0.52355345,0.42556711){\color[rgb]{0,0,0}\makebox(0,0)[lt]{\lineheight{1.25}\smash{\begin{tabular}[t]{l}$z$\end{tabular}}}}%
    \put(0.08293708,0.37632841){\color[rgb]{0,0,0}\makebox(0,0)[lt]{\lineheight{1.25}\smash{\begin{tabular}[t]{l}$\vec{E^i}$\end{tabular}}}}%
    \put(0,0){\includegraphics[width=\unitlength,page=2]{TE_TM_inc_parallel.pdf}}%
    \put(0.90747746,0.25443035){\color[rgb]{0,0,0}\makebox(0,0)[lt]{\lineheight{1.25}\smash{\begin{tabular}[t]{l}$\vec{E^s}$\end{tabular}}}}%
    \put(0,0){\includegraphics[width=\unitlength,page=3]{TE_TM_inc_parallel.pdf}}%
    \put(0.37429167,0.19480766){\color[rgb]{0,0,0}\makebox(0,0)[lt]{\lineheight{1.25}\smash{\begin{tabular}[t]{l}$\theta^i$\end{tabular}}}}%
    \put(0.04093064,0.23671957){\color[rgb]{0,0,0}\makebox(0,0)[lt]{\lineheight{1.25}\smash{\begin{tabular}[t]{l}$\vec{H^i}$\end{tabular}}}}%
    \put(0.78282972,0.32962719){\color[rgb]{0,0,0}\makebox(0,0)[lt]{\lineheight{1.25}\smash{\begin{tabular}[t]{l}$\vec{H^s}$\end{tabular}}}}%
    \put(0,0){\includegraphics[width=\unitlength,page=4]{TE_TM_inc_parallel.pdf}}%
    \put(0.94192022,0.10441079){\color[rgb]{0,0,0}\makebox(0,0)[lt]{\lineheight{1.25}\smash{\begin{tabular}[t]{l}$x$\end{tabular}}}}%
    \put(0,0){\includegraphics[width=\unitlength,page=5]{TE_TM_inc_parallel.pdf}}%
    \put(0.56077821,0.19356639){\color[rgb]{0,0,0}\makebox(0,0)[lt]{\lineheight{1.25}\smash{\begin{tabular}[t]{l}$\theta^s$\end{tabular}}}}%
    \put(0,0){\includegraphics[width=\unitlength,page=6]{TE_TM_inc_parallel.pdf}}%
    \put(0.08428058,0.14851626){\color[rgb]{0,0,0}\makebox(0,0)[lt]{\lineheight{1.25}\smash{\begin{tabular}[t]{l}$D_x$\end{tabular}}}}%
    \put(0,0){\includegraphics[width=\unitlength,page=7]{TE_TM_inc_parallel.pdf}}%
    \put(0.27480644,0.15099797){\color[rgb]{0,0,0}\makebox(0,0)[lt]{\lineheight{1.25}\smash{\begin{tabular}[t]{l}$w_x$\end{tabular}}}}%
    \put(0,0){\includegraphics[width=\unitlength,page=8]{TE_TM_inc_parallel.pdf}}%
  \end{picture}%
\endgroup%

%% file: Figure/TE_TM_inc/TE_TM_inc_hortogonal.pdf_tex
\begingroup%
  \makeatletter%
  \providecommand\color[2][]{%
    \errmessage{(Inkscape) Color is used for the text in Inkscape, but the package 'color.sty' is not loaded}%
    \renewcommand\color[2][]{}%
  }%
  \providecommand\transparent[1]{%
    \errmessage{(Inkscape) Transparency is used (non-zero) for the text in Inkscape, but the package 'transparent.sty' is not loaded}%
    \renewcommand\transparent[1]{}%
  }%
  \providecommand\rotatebox[2]{#2}%
  \newcommand*\fsize{\dimexpr\f@size pt\relax}%
  \newcommand*\lineheight[1]{\fontsize{\fsize}{#1\fsize}\selectfont}%
  \ifx\svgwidth\undefined%
    \setlength{\unitlength}{165.2703509bp}%
    \ifx\svgscale\undefined%
      \relax%
    \else%
      \setlength{\unitlength}{\unitlength * \real{\svgscale}}%
    \fi%
  \else%
    \setlength{\unitlength}{\svgwidth}%
  \fi%
  \global\let\svgwidth\undefined%
  \global\let\svgscale\undefined%
  \makeatother%
  \begin{picture}(1,0.45921105)%
    \lineheight{1}%
    \setlength\tabcolsep{0pt}%
    \put(0,0){\includegraphics[width=\unitlength,page=1]{TE_TM_inc_hortogonal.pdf}}%
    \put(0.22454515,0.3518438){\color[rgb]{0,0,0}\makebox(0,0)[lt]{\lineheight{1.25}\smash{\begin{tabular}[t]{l}$\vec{H^i_z}$\end{tabular}}}}%
    \put(0.21903957,0.26241375){\color[rgb]{0,0,0}\makebox(0,0)[lt]{\lineheight{1.25}\smash{\begin{tabular}[t]{l}$\vec{H^i_x}$\end{tabular}}}}%
    \put(0.52355345,0.42556711){\color[rgb]{0,0,0}\makebox(0,0)[lt]{\lineheight{1.25}\smash{\begin{tabular}[t]{l}$z$\end{tabular}}}}%
    \put(0.09055604,0.38728416){\color[rgb]{0,0,0}\makebox(0,0)[lt]{\lineheight{1.25}\smash{\begin{tabular}[t]{l}$\vec{H^i}$\end{tabular}}}}%
    \put(0,0){\includegraphics[width=\unitlength,page=2]{TE_TM_inc_hortogonal.pdf}}%
    \put(0.82168037,0.20039955){\color[rgb]{0,0,0}\makebox(0,0)[lt]{\lineheight{1.25}\smash{\begin{tabular}[t]{l}$\vec{E^s}$\end{tabular}}}}%
    \put(0,0){\includegraphics[width=\unitlength,page=3]{TE_TM_inc_hortogonal.pdf}}%
    \put(0.37429167,0.19480766){\color[rgb]{0,0,0}\makebox(0,0)[lt]{\lineheight{1.25}\smash{\begin{tabular}[t]{l}$\theta^i$\end{tabular}}}}%
    \put(0.04651074,0.23905739){\color[rgb]{0,0,0}\makebox(0,0)[lt]{\lineheight{1.25}\smash{\begin{tabular}[t]{l}$\vec{E^i}$\end{tabular}}}}%
    \put(0.75544952,0.35917368){\color[rgb]{0,0,0}\makebox(0,0)[lt]{\lineheight{1.25}\smash{\begin{tabular}[t]{l}$\vec{H^s}$\end{tabular}}}}%
    \put(0,0){\includegraphics[width=\unitlength,page=4]{TE_TM_inc_hortogonal.pdf}}%
    \put(0.94192022,0.10441079){\color[rgb]{0,0,0}\makebox(0,0)[lt]{\lineheight{1.25}\smash{\begin{tabular}[t]{l}$x$\end{tabular}}}}%
    \put(0,0){\includegraphics[width=\unitlength,page=5]{TE_TM_inc_hortogonal.pdf}}%
    \put(0.56215965,0.19362796){\color[rgb]{0,0,0}\makebox(0,0)[lt]{\lineheight{1.25}\smash{\begin{tabular}[t]{l}$\theta^s$\end{tabular}}}}%
    \put(0,0){\includegraphics[width=\unitlength,page=6]{TE_TM_inc_hortogonal.pdf}}%
    \put(0.08473439,0.14828935){\color[rgb]{0,0,0}\makebox(0,0)[lt]{\lineheight{1.25}\smash{\begin{tabular}[t]{l}$D_x$\end{tabular}}}}%
    \put(0,0){\includegraphics[width=\unitlength,page=7]{TE_TM_inc_hortogonal.pdf}}%
    \put(0.27526025,0.15077107){\color[rgb]{0,0,0}\makebox(0,0)[lt]{\lineheight{1.25}\smash{\begin{tabular}[t]{l}$w_x$\end{tabular}}}}%
    \put(0,0){\includegraphics[width=\unitlength,page=8]{TE_TM_inc_hortogonal.pdf}}%
  \end{picture}%
\endgroup%

%% file: Figure/RIS_geometry/RIS_geometry_final.pdf_tex
\begingroup%
  \makeatletter%
  \providecommand\color[2][]{%
    \errmessage{(Inkscape) Color is used for the text in Inkscape, but the package 'color.sty' is not loaded}%
    \renewcommand\color[2][]{}%
  }%
  \providecommand\transparent[1]{%
    \errmessage{(Inkscape) Transparency is used (non-zero) for the text in Inkscape, but the package 'transparent.sty' is not loaded}%
    \renewcommand\transparent[1]{}%
  }%
  \providecommand\rotatebox[2]{#2}%
  \newcommand*\fsize{\dimexpr\f@size pt\relax}%
  \newcommand*\lineheight[1]{\fontsize{\fsize}{#1\fsize}\selectfont}%
  \ifx\svgwidth\undefined%
    \setlength{\unitlength}{700.3616357bp}%
    \ifx\svgscale\undefined%
      \relax%
    \else%
      \setlength{\unitlength}{\unitlength * \real{\svgscale}}%
    \fi%
  \else%
    \setlength{\unitlength}{\svgwidth}%
  \fi%
  \global\let\svgwidth\undefined%
  \global\let\svgscale\undefined%
  \makeatother%
  \begin{picture}(1,0.3237657)%
    \lineheight{1}%
    \setlength\tabcolsep{0pt}%
    \put(0,0){\includegraphics[width=\unitlength,page=1]{RIS_geometry_final.pdf}}%
    \put(0.04605183,0.23340167){\color[rgb]{0,0,0}\makebox(0,0)[lt]{\lineheight{1.25}\smash{\begin{tabular}[t]{l}$TX$\end{tabular}}}}%
    \put(-0.00106669,0.21336102){\color[rgb]{0,0,0}\makebox(0,0)[lt]{\lineheight{1.25}\smash{\begin{tabular}[t]{l}$(x_t, y_t, z_t)$\end{tabular}}}}%
    \put(0.51809948,0.26458246){\color[rgb]{0,0,0}\makebox(0,0)[lt]{\lineheight{1.25}\smash{\begin{tabular}[t]{l}$RX$\end{tabular}}}}%
    \put(0.49775284,0.24836636){\color[rgb]{0,0,0}\makebox(0,0)[lt]{\lineheight{1.25}\smash{\begin{tabular}[t]{l}$(x_r, y_r, z_r)$\end{tabular}}}}%
    \put(0,0){\includegraphics[width=\unitlength,page=2]{RIS_geometry_final.pdf}}%
    \put(0.14991711,0.16600237){\color[rgb]{0,0,0}\rotatebox{-50.699074}{\makebox(0,0)[lt]{\lineheight{1.25}\smash{\begin{tabular}[t]{l}$r^t_{n,m}$\end{tabular}}}}}%
    \put(0.42974114,0.18165273){\color[rgb]{0,0,0}\rotatebox{51.897069}{\makebox(0,0)[lt]{\lineheight{1.25}\smash{\begin{tabular}[t]{l}$r^r_{n,m}$\end{tabular}}}}}%
    \put(0.47742848,0.08236742){\color[rgb]{0,0,0}\makebox(0,0)[lt]{\lineheight{1.25}\smash{\begin{tabular}[t]{l}$x$\end{tabular}}}}%
    \put(0.34749466,0.20955151){\color[rgb]{0,0,0}\makebox(0,0)[lt]{\lineheight{1.25}\smash{\begin{tabular}[t]{l}$y$\end{tabular}}}}%
    \put(0.30025774,0.21654958){\color[rgb]{0,0,0}\makebox(0,0)[lt]{\lineheight{1.25}\smash{\begin{tabular}[t]{l}$z$\end{tabular}}}}%
    \put(0,0){\includegraphics[width=\unitlength,page=3]{RIS_geometry_final.pdf}}%
    \put(0.41413507,0.10209619){\color[rgb]{0,0,0}\makebox(0,0)[lt]{\lineheight{1.25}\smash{\begin{tabular}[t]{l}$\varphi^r$\end{tabular}}}}%
    \put(0,0){\includegraphics[width=\unitlength,page=4]{RIS_geometry_final.pdf}}%
    \put(0.16354206,0.20799097){\color[rgb]{0,0,0}\makebox(0,0)[lt]{\lineheight{1.25}\smash{\begin{tabular}[t]{l}$\theta^{tx}_{n,m}$\end{tabular}}}}%
    \put(0,0){\includegraphics[width=\unitlength,page=5]{RIS_geometry_final.pdf}}%
    \put(0.39474751,0.25445072){\color[rgb]{0,0,0}\makebox(0,0)[lt]{\lineheight{1.25}\smash{\begin{tabular}[t]{l}$\theta^{rx}_{n,m}$\end{tabular}}}}%
    \put(0.29983923,0.12559776){\color[rgb]{0,0,0}\makebox(0,0)[lt]{\lineheight{1.25}\smash{\begin{tabular}[t]{l}$\theta^r$\end{tabular}}}}%
    \put(0,0){\includegraphics[width=\unitlength,page=6]{RIS_geometry_final.pdf}}%
    \put(0.26609802,0.12454314){\color[rgb]{0,0,0}\makebox(0,0)[lt]{\lineheight{1.25}\smash{\begin{tabular}[t]{l}$\theta^t$\end{tabular}}}}%
    \put(0.19584175,0.18599316){\color[rgb]{0,0,0}\rotatebox{-38.777588}{\makebox(0,0)[lt]{\lineheight{1.25}\smash{\begin{tabular}[t]{l}$r^t$\end{tabular}}}}}%
    \put(0.39367472,0.20465312){\color[rgb]{0,0,0}\rotatebox{47.371594}{\makebox(0,0)[lt]{\lineheight{1.25}\smash{\begin{tabular}[t]{l}$r^r$\end{tabular}}}}}%
    \put(0,0){\includegraphics[width=\unitlength,page=7]{RIS_geometry_final.pdf}}%
    \put(0.32137635,0.10513724){\color[rgb]{0,0,0}\makebox(0,0)[lt]{\lineheight{1.25}\smash{\begin{tabular}[t]{l}$\varphi^t$\end{tabular}}}}%
    \put(0,0){\includegraphics[width=\unitlength,page=8]{RIS_geometry_final.pdf}}%
    \put(0.90487401,0.02278214){\color[rgb]{0,0,0}\makebox(0,0)[lt]{\lineheight{1.25}\smash{\begin{tabular}[t]{l}$x^\prime$\end{tabular}}}}%
    \put(0.73620874,0.19288818){\color[rgb]{0,0,0}\makebox(0,0)[lt]{\lineheight{1.25}\smash{\begin{tabular}[t]{l}$z^\prime$\end{tabular}}}}%
    \put(0,0){\includegraphics[width=\unitlength,page=9]{RIS_geometry_final.pdf}}%
    \put(0.86768158,0.07149265){\color[rgb]{0,0,0}\makebox(0,0)[lt]{\lineheight{1.25}\smash{\begin{tabular}[t]{l}$\varphi^r_{n,m}$\end{tabular}}}}%
    \put(0,0){\includegraphics[width=\unitlength,page=10]{RIS_geometry_final.pdf}}%
    \put(0.73640865,0.11234387){\color[rgb]{0,0,0}\makebox(0,0)[lt]{\lineheight{1.25}\smash{\begin{tabular}[t]{l}$\theta^r_{n,m}$\end{tabular}}}}%
    \put(0.6837406,0.10120903){\color[rgb]{0,0,0}\makebox(0,0)[lt]{\lineheight{1.25}\smash{\begin{tabular}[t]{l}$\theta^t_{n,m}$\end{tabular}}}}%
    \put(0.78929947,0.05050453){\color[rgb]{0,0,0}\makebox(0,0)[lt]{\lineheight{1.25}\smash{\begin{tabular}[t]{l}$\varphi^t_{n,m}$\end{tabular}}}}%
    \put(0,0){\includegraphics[width=\unitlength,page=11]{RIS_geometry_final.pdf}}%
    \put(0.31107072,0.31725678){\color[rgb]{0,0,0}\makebox(0,0)[t]{\lineheight{1.25}\smash{\begin{tabular}[t]{c}Blocking Object\end{tabular}}}}%
    \put(0,0){\includegraphics[width=\unitlength,page=12]{RIS_geometry_final.pdf}}%
    \put(0.13083499,0.00226802){\color[rgb]{0,0,0}\makebox(0,0)[lt]{\lineheight{1.25}\smash{\begin{tabular}[t]{l}$D_x$\end{tabular}}}}%
    \put(0.10167004,0.03286372){\color[rgb]{0,0,0}\makebox(0,0)[lt]{\lineheight{1.25}\smash{\begin{tabular}[t]{l}$D_y$\end{tabular}}}}%
    \put(0,0){\includegraphics[width=\unitlength,page=13]{RIS_geometry_final.pdf}}%
    \put(0.27754016,0.05634735){\color[rgb]{0,0,0}\makebox(0,0)[lt]{\lineheight{1.25}\smash{\begin{tabular}[t]{l}$z^\prime$\end{tabular}}}}%
    \put(0.31487439,0.02604552){\color[rgb]{0,0,0}\makebox(0,0)[lt]{\lineheight{1.25}\smash{\begin{tabular}[t]{l}$x^\prime$\end{tabular}}}}%
  \end{picture}%
\endgroup%

%% file: Figure/TE_TM_polarization/floquet_v2.pdf_tex
\begingroup%
  \makeatletter%
  \providecommand\color[2][]{%
    \errmessage{(Inkscape) Color is used for the text in Inkscape, but the package 'color.sty' is not loaded}%
    \renewcommand\color[2][]{}%
  }%
  \providecommand\transparent[1]{%
    \errmessage{(Inkscape) Transparency is used (non-zero) for the text in Inkscape, but the package 'transparent.sty' is not loaded}%
    \renewcommand\transparent[1]{}%
  }%
  \providecommand\rotatebox[2]{#2}%
  \newcommand*\fsize{\dimexpr\f@size pt\relax}%
  \newcommand*\lineheight[1]{\fontsize{\fsize}{#1\fsize}\selectfont}%
  \ifx\svgwidth\undefined%
    \setlength{\unitlength}{167.67524299bp}%
    \ifx\svgscale\undefined%
      \relax%
    \else%
      \setlength{\unitlength}{\unitlength * \real{\svgscale}}%
    \fi%
  \else%
    \setlength{\unitlength}{\svgwidth}%
  \fi%
  \global\let\svgwidth\undefined%
  \global\let\svgscale\undefined%
  \makeatother%
  \begin{picture}(1,0.32547628)%
    \lineheight{1}%
    \setlength\tabcolsep{0pt}%
    \put(0,0){\includegraphics[width=\unitlength,page=1]{floquet_v2.pdf}}%
    \put(0.08864149,0.30252919){\color[rgb]{0,0,0}\makebox(0,0)[lt]{\lineheight{1.25}\smash{\begin{tabular}[t]{l}$z$\end{tabular}}}}%
    \put(0,0){\includegraphics[width=\unitlength,page=2]{floquet_v2.pdf}}%
    \put(0.94650531,0.01386485){\color[rgb]{0,0,0}\makebox(0,0)[lt]{\lineheight{1.25}\smash{\begin{tabular}[t]{l}$x$\end{tabular}}}}%
    \put(0,0){\includegraphics[width=\unitlength,page=3]{floquet_v2.pdf}}%
    \put(0.13386691,0.19251209){\color[rgb]{0,0,0}\makebox(0,0)[lt]{\lineheight{1.25}\smash{\begin{tabular}[t]{l}$D_x$\end{tabular}}}}%
  \end{picture}%
\endgroup%

%% file: Figure/Impedance_surf_scattering/frestanding_metasurface_geometry.pdf_tex
\begingroup%
  \makeatletter%
  \providecommand\color[2][]{%
    \errmessage{(Inkscape) Color is used for the text in Inkscape, but the package 'color.sty' is not loaded}%
    \renewcommand\color[2][]{}%
  }%
  \providecommand\transparent[1]{%
    \errmessage{(Inkscape) Transparency is used (non-zero) for the text in Inkscape, but the package 'transparent.sty' is not loaded}%
    \renewcommand\transparent[1]{}%
  }%
  \providecommand\rotatebox[2]{#2}%
  \newcommand*\fsize{\dimexpr\f@size pt\relax}%
  \newcommand*\lineheight[1]{\fontsize{\fsize}{#1\fsize}\selectfont}%
  \ifx\svgwidth\undefined%
    \setlength{\unitlength}{182.7110711bp}%
    \ifx\svgscale\undefined%
      \relax%
    \else%
      \setlength{\unitlength}{\unitlength * \real{\svgscale}}%
    \fi%
  \else%
    \setlength{\unitlength}{\svgwidth}%
  \fi%
  \global\let\svgwidth\undefined%
  \global\let\svgscale\undefined%
  \makeatother%
  \begin{picture}(1,0.52794644)%
    \lineheight{1}%
    \setlength\tabcolsep{0pt}%
    \put(0,0){\includegraphics[width=\unitlength,page=1]{frestanding_metasurface_geometry.pdf}}%
    \put(0.17027238,0.46366668){\color[rgb]{0,0,0}\makebox(0,0)[lt]{\lineheight{1.25}\smash{\begin{tabular}[t]{l}$\vec{E}^i_z$\end{tabular}}}}%
    \put(0.16529234,0.38277317){\color[rgb]{0,0,0}\makebox(0,0)[lt]{\lineheight{1.25}\smash{\begin{tabular}[t]{l}$\vec{E}^i_y$\end{tabular}}}}%
    \put(0.47357755,0.49751399){\color[rgb]{0,0,0}\makebox(0,0)[lt]{\lineheight{1.25}\smash{\begin{tabular}[t]{l}$z$\end{tabular}}}}%
    \put(0.04218154,0.48581411){\color[rgb]{0,0,0}\makebox(0,0)[lt]{\lineheight{1.25}\smash{\begin{tabular}[t]{l}$\vec{E}^i$\end{tabular}}}}%
    \put(0,0){\includegraphics[width=\unitlength,page=2]{frestanding_metasurface_geometry.pdf}}%
    \put(0.82085403,0.32629377){\color[rgb]{0,0,0}\makebox(0,0)[lt]{\lineheight{1.25}\smash{\begin{tabular}[t]{l}$\vec{E}^s$\end{tabular}}}}%
    \put(0,0){\includegraphics[width=\unitlength,page=3]{frestanding_metasurface_geometry.pdf}}%
    \put(0.41031657,0.30088998){\color[rgb]{0,0,0}\makebox(0,0)[lt]{\lineheight{1.25}\smash{\begin{tabular}[t]{l}$\theta^i$\end{tabular}}}}%
    \put(0.00418484,0.35953163){\color[rgb]{0,0,0}\makebox(0,0)[lt]{\lineheight{1.25}\smash{\begin{tabular}[t]{l}$\vec{H}^i$\end{tabular}}}}%
    \put(0.70810456,0.41073203){\color[rgb]{0,0,0}\makebox(0,0)[lt]{\lineheight{1.25}\smash{\begin{tabular}[t]{l}$\vec{H}^s$\end{tabular}}}}%
    \put(0,0){\includegraphics[width=\unitlength,page=4]{frestanding_metasurface_geometry.pdf}}%
    \put(0.85200905,0.2070137){\color[rgb]{0,0,0}\makebox(0,0)[lt]{\lineheight{1.25}\smash{\begin{tabular}[t]{l}$y$\end{tabular}}}}%
    \put(0,0){\includegraphics[width=\unitlength,page=5]{frestanding_metasurface_geometry.pdf}}%
    \put(0.50724901,0.28765895){\color[rgb]{0,0,0}\makebox(0,0)[lt]{\lineheight{1.25}\smash{\begin{tabular}[t]{l}$\theta^s$\end{tabular}}}}%
    \put(0,0){\includegraphics[width=\unitlength,page=6]{frestanding_metasurface_geometry.pdf}}%
    \put(0.23848127,0.2131232){\color[rgb]{0,0,0}\makebox(0,0)[lt]{\lineheight{1.25}\smash{\begin{tabular}[t]{l}$\vec{H}^+$\end{tabular}}}}%
    \put(0,0){\includegraphics[width=\unitlength,page=7]{frestanding_metasurface_geometry.pdf}}%
    \put(0.575784,0.01267262){\color[rgb]{0,0,0}\makebox(0,0)[lt]{\lineheight{1.25}\smash{\begin{tabular}[t]{l}$\vec{H}^t$\end{tabular}}}}%
    \put(0.70887003,0.0622552){\color[rgb]{0,0,0}\makebox(0,0)[lt]{\lineheight{1.25}\smash{\begin{tabular}[t]{l}$\vec{E}^t$\end{tabular}}}}%
    \put(0.09575777,0.13061379){\color[rgb]{0,0,0}\makebox(0,0)[lt]{\lineheight{1.25}\smash{\begin{tabular}[t]{l}$J_{av}$\end{tabular}}}}%
    \put(0,0){\includegraphics[width=\unitlength,page=8]{frestanding_metasurface_geometry.pdf}}%
    \put(0.31190207,0.30252761){\color[rgb]{0,0,0}\makebox(0,0)[lt]{\lineheight{1.25}\smash{\begin{tabular}[t]{l}$\vec{E^+}$\end{tabular}}}}%
    \put(0,0){\includegraphics[width=\unitlength,page=9]{frestanding_metasurface_geometry.pdf}}%
    \put(0.38522942,0.1380046){\color[rgb]{0,0,0}\makebox(0,0)[lt]{\lineheight{1.25}\smash{\begin{tabular}[t]{l}$\vec{H}^-$\end{tabular}}}}%
    \put(0,0){\includegraphics[width=\unitlength,page=10]{frestanding_metasurface_geometry.pdf}}%
    \put(0.53561594,0.21304203){\color[rgb]{0,0,0}\makebox(0,0)[lt]{\lineheight{1.25}\smash{\begin{tabular}[t]{l}$\vec{E^-}$\end{tabular}}}}%
  \end{picture}%
\endgroup%